\documentclass[a4paper,fleqn]{cas-dc}

\usepackage[square,sort,comma,numbers]{natbib}
\usepackage{amsfonts}
\usepackage{booktabs}
\usepackage{smartdiagram}
\usesmartdiagramlibrary{additions}
\usepackage{tikz}
\usetikzlibrary{arrows.meta}
\usetikzlibrary{plotmarks}

\def\tsc#1{\csdef{#1}{\textsc{\lowercase{#1}}\xspace}}
\tsc{WGM}
\tsc{QE}
\def\myref{\text{\scriptsize ref}}
\newcommand\Tstrut{\rule{0.pt}{2.6ex}} 

\newcommand\eq{\textrm{eq}}
\newcommand{\vek}[2][0]{
    \ifnum0=#1
    \else
        \begingroup
        \renewcommand*{\arraystretch}{#1}
    \fi
    \begin{pmatrix}
        #2
    \end{pmatrix}
    \ifnum0=#1
    \else
        \endgroup
    \fi
}

\begin{document}

\let\WriteBookmarks\relax
\def\floatpagepagefraction{1}
\def\textpagefraction{.001}

\shorttitle{CBC for thermal LBM}    

\shortauthors{F. Klass, A. Gabbana, A. Bartel}  

\title [mode = title]{Characteristic Boundary Condition for Thermal Lattice Boltzmann Methods}  

\author[1]{Friedemann Klass}[orcid=0000-0002-8566-0918]

\ead{klass@math.uni-wuppertal.de}

\author[2]{Alessandro Gabbana}[orcid=0000-0002-8367-6596]
\author[1]{Andreas Bartel}[orcid=0000-0003-1979-179X]

\affiliation[1]{organization={University of Wuppertal},
            addressline={Gaussstrasse 20}, 
            city={Wuppertal},
            postcode={42119},
            country={Germany}}

\affiliation[2]{organization={Eindhoven University of Technology},
            city={Eindhoven},
            postcode={5600 MB},
            country={The Netherlands}}

\cortext[1]{Corresponding author}

\begin{abstract}
    We introduce a non-reflecting boundary condition for the simulation 
    of thermal flows with the lattice Boltzmann Method (LBM).
    We base the derivation on the locally one-dimensional inviscid analysis,
    and define target macroscopic values at the boundary
    aiming at minimizing the effect of reflections of outgoing waves
    on the bulk dynamics.
    The resulting macroscopic target values are then enforced in the LBM 
    using a mesoscopic Dirichlet boundary condition.
    We present a procedure which allows to implement the boundary 
    treatment for both single-speed and high order multi-speed LBM models,
    by conducting a layerwise characteristic analysis.
    We demonstrate the effectiveness of our approach by providing
    qualitative and quantitative comparison of
    several strategies for the implementation of a open boundary condition
    in standard numerical benchmarks.
    We show that our approach allows to achieve increasingly high accuracy by
    relaxing transversal and viscous terms towards prescribed target values.
\end{abstract}

\begin{keywords}
	  CFD \sep
    Non-reflective Boundary Conditions \sep Characteristic Boundary Condition 
    \sep  Locally One-Dimensional Inviscid \sep Thermal flows \sep  High order lattice Boltzmann methods
\end{keywords}

\maketitle
%
\section{Introduction}\label{sec:intro}
%
In many numerical simulations,  the physical domain of a given problem is infeasibly large and only 
information about a small region of interest, encapsulated in the physical domain, is required. 
In these situations, the computational domain can be obtained by truncating the original domain. 
This gives rise to artificial boundaries that need to be treated using a boundary condition (BC).
Since these artificial boundaries should not interact with the bulk dynamics, it is beneficial 
to employ non-reflecting boundary conditions (NRBC), which let waves crossing the boundary out of the 
computational domain without causing reflection effects.

The so-called characteristic boundary condition (CBC) is among the most popular choices for NRBC. 
Its working principle consists of identifying incoming and outgoing waves at the boundary, to then
modulate amplitude variations of incoming waves. 
The CBC was originally developed for nonlinear hyperbolic systems \cite{hedstrom-jocp-1979,thompson-jocp-1987} 
and later extended to the Navier-Stokes equations \cite{poinsot-jocp-1992}, where it is widely used 
in the area of computational aero-acoustics \cite{kim-aiaa-2000,sandberg-aiaa-2006}. 
Furthermore, CBCs have been successfully applied to magneto-hydrodynamics \cite{wu-cmame-1987,cimino-cf-2016}, 
reacting \cite{yoo-ctan-2007} and turbulent flow in two \cite{yoo-ctan-2005}  
and three \cite{lodato-jocp-2008} spatial dimensions.

In this work, we present the derivation of a CBC for thermal compressible flows, modeled by the 
Navier-Stokes-Fourier equations. We employ the lattice Boltzmann method (LBM) for the time evolution
of the fluid bulk dynamic.
The LBM is an established algorithm for the simulation of fluid flows that can be derived as a 
systematic approximation of the Boltzmann equation \cite{shan-jofm-2006,arumugaperumal-aej-2015}.
It has gained a lot of popularity due to its simple algorithmic structure, which makes it highly 
amenable to large scale parallelization~\cite{calore-pm-2016,bauer-cmwa-2021,latt-plos-2021}, the 
ability to handle complex geometries \cite{guo-pre-2002} as well as multi-phase \cite{chen-ijhmt-2014} 
and multi-component \cite{martys-pre-1996} flows. 
The LBM provides the description of a fluid flow in terms of a discrete 
set of particle distribution functions (populations) sitting at the sites of a discrete lattice, 
with the macroscopic behavior emerging from the velocity moments of the distribution.
Continuous efforts are being made to extend
the range of applicability of LBM and tackle problems such as
thermal compressible flows. There are three main approaches 
to thermal LBM present in literature: i) hybrid coupling with a macroscopic solver 
(e.g. finite differences, or finite volume) for evolving the energy equation~\cite{lallemand-ijmp-2003}, 
ii) the double distribution approach, where a second set of populations is used to evolve 
the temperature field~\cite{li-pre-2012} and iii) models based on high order quadrature rules~\cite{shan-jofm-2006,shan-pre-2010}.
The latter approach provides an elegant and self-consistent kinetic description of 
thermal compressible flow  \cite{philippi-pre-2006,scagliarini-pof-2010} via the high order moments
of the particle distribution function.
However, higher order models give rise to multi-speed velocity stencils, i.e., 
discrete velocity stencils with a maximum displacement greater than one, which generally complicate
the definition of boundary conditions.

Few implementations of characteristic boundary conditions for isothermal (single-speed) LBM have 
been reported in the literature \cite{izquierdo-aps-2008,dehee-cps-2008,heubes-jcam-2014,jung-jocp-2015},
finding application in the simulation of fluid flows in high Reynolds number regimes \cite{wissocq-jocp-2017,
jacob-pof-2019,gianoli-ijnmf-2022}.
To the best of our knowledge, there are currently only two works in the literature where the application 
of characteristic BCs for multi-speed LBM is discussed:
the CBC for the Navier-Stokes equations presented in \cite{izquierdo-aps-2008} has been used in 
conjunction with multi-speed velocity stencils in a previous article~\cite{klass-cicp-2023} 
by the authors of the work at hand, and recently multi-speed CBC was used 
in the context of thermo-acoustic problems governed by the Euler equations~\cite{chen-jocp-2023}.

In the work at hand, we develop a CBC suitable for thermal compressible flows governed 
by the Navier-Stokes-Fourier equations. 
We will focus our analysis on its coupling with multi-speed LBM, however we shall remark 
that the procedure here described is general and can be employed also in combination to the 
other approaches for thermal LBM listed above. 

This article is organized as follows:
Sec.~\ref{sec:lbm} provides a brief description of the LBM scheme used in this work. 
In Sec.~\ref{sec:cbc}, we present the derivation of a CBC for the Navier-Stokes-Fourier equations 
as well as a simplified version where transversal and viscous terms are discarded. 
We then provide details on how to couple the macroscopic BC with the mesoscopic layer.
Subsequently, in Sec.~\ref{sec:numerics}, we benchmark the numerical accuracy of the BC 
in several test-beds, comparing different realization of the CBC with 
a simple zero-gradient extrapolation.
Finally, concluding remarks and future directions are summarized in Sec.~\ref{sec:conclusions}.

\section{Thermal Lattice Boltzmann Method}\label{sec:lbm}

In this section, we provide a succinct overview of the 
LBM which we use in this work to solve the Navier-Stokes-Fourier equations.
The reader not familiar with LBM is referred to Ref.~\cite{kruger-book-2017,
succi-book-2018} for a more comprehensive introduction.
We remark that while we focus on $d=2$ spatial dimensions during this work, 
to ease the presentation and provide a broad and comprehensive picture of the behavior of CBCs, 
the generalization to three spatial dimensions is conceptually straightforward (see Appendix A).

The LBM operates at the mesoscopic level and provides the time evolution 
of a fluid flow via the synthetic dynamics of a set of discrete velocity
distribution functions, governed by the discrete lattice Boltzmann equation:
(for $i = 1,\ \dots,\ q$)
\begin{equation}\label{eq:lbe}
  f_i(\mathbf{x}+ \mathbf{c}_{i} \Delta t , t + \Delta t) 
  =  
  f_i(\mathbf{x},t) + \Omega_i(\mathbf{x},t) 
  .
\end{equation} 
In the above, $f_i(\mathbf{x}, t)$ are the discrete single particle distribution
functions (to which we will refer to as lattice populations), 
defined at each node $\mathbf{x}$ of a Cartesian grid, and corresponding to
each of the $q$ components of a velocity stencil 
$\{ \mathbf{c}_i = (c_{i,x},c_{i,y}):\, i = 1,\, \dots,\, q\}$,
while $\Omega_i$ is the collision operator. 
Hereafter, we work in rescaled units such that the Boltzmann constant $k_b$ and the molecular mass $m$ of the gas 
are set to unity~\cite{scagliarini-pof-2010}.

Hydrodynamic quantities of interest, such as density $\rho$, 
velocity $\mathbf{u}=(u_x,u_y)^\top$ and internal energy density $\rho e$ can be calculated from the 
velocity moments of the distribution: 
\begin{flalign}\label{eq:macro}
  \rho            = \sum\limits_{i=1}^q f_i                                             ,  \;     
  \rho \mathbf{u} = \sum\limits_{i=1}^q f_i \mathbf{c}_{i}                              ,  \;
  2 \rho e        = \sum\limits_{i=1}^q f_i \vert \mathbf{c}_{i} - \mathbf{u}\vert^2    .
  &&
\end{flalign}
For a monoatomic ideal gas, the link between temperature and internal energy is given by~\cite{shan-jofm-2006}
\begin{equation}
    e = \frac{d}{2} T .
\end{equation}
\begin{figure*}
    \centering
    \begin{tikzpicture}[scale = 0.8]

        \foreach \y in {-3, ...,3} {
            \foreach \x in {-3,...,10}{
                \draw
                (\x,\y) circle(2pt);
            }
        }

        \foreach \x in {-3,3} {
                
            \draw[-{Latex[width=2.2mm,length=2.7mm]},semithick]   (0,0) -- (\x , 0);
            \draw[-{Latex[width=2.2mm,length=2.7mm]},semithick]   (0,0) -- (0  , \x );
            
        }
        \foreach \x in {-2,2} {
            
            \foreach \y in {-2,2} {
               \draw[-{Latex[width=2.2mm,length=2.7mm]},semithick]   (0,0) -- (\x , \y);
            }
        }
        \foreach \x in {-1,1} {
            
            \draw[-{Latex[width=2.2mm,length=2.7mm]},semithick]   (0,0) -- (\x ,  0);
            \draw[-{Latex[width=2.2mm,length=2.7mm]},semithick]   (0,0) -- (0  , \x);

            \foreach \y in {-1,1} {
                \draw[-{Latex[width=2.2mm,length=2.7mm]},semithick]   (0,0) -- (\x , \y);
            }
        }

        \foreach \y in {1,2,3}{
            \foreach \x in {-\y,\y} {
                    
                \draw[-{Latex[width=2.2mm,length=2.7mm]},semithick]   (7,0) -- (7+\x , 0);
                \draw[-{Latex[width=2.2mm,length=2.7mm]},semithick]   (7,0) -- (7+0  , \x );
                
            }
        }
        \foreach \z in {1,2}{
            \foreach \x in {-\z,\z} {
                
                \foreach \y in {-\z,\z} {
                   \draw[-{Latex[width=2.2mm,length=2.7mm]},semithick]   (7,0) -- (7+\x , \y);
                }
            }
        }
        \foreach \z in {2,3}{
            \foreach \x in {-1,1} {
                
                \foreach \y in {-\z,\z} {
                   \draw[-{Latex[width=2.2mm,length=2.7mm]},semithick]   (7,0) -- (7+\x , \y);
                   \draw[-{Latex[width=2.2mm,length=2.7mm]},semithick]   (7,0) -- (7+\y , \x);
                }
            }
        }

        \draw[-{Latex[width=2.2mm,length=2.7mm]},semithick]   (-6,-3) -- (-4 , -3) node[below]{x};
        \draw[-{Latex[width=2.2mm,length=2.7mm]},semithick]   (-6,-3) -- (-6 , -1) node[left]{y};
    \end{tikzpicture}
    \caption{Schematic representation of the velocity directions for the D2Q17 
             (left) and D2Q37 (right) velocity stencils.
            }\label{fig1:stencils}
\end{figure*}
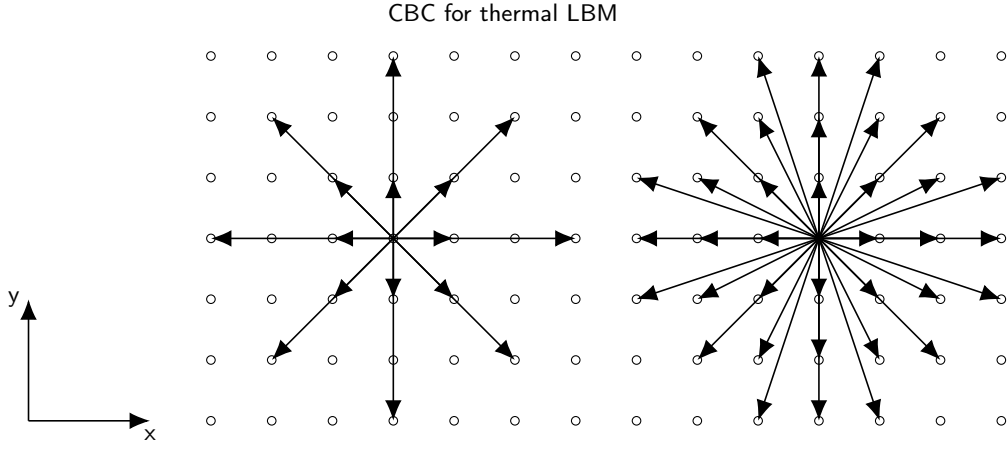

In Eq.~\eqref{eq:macro}, equality holds if the lattice velocities 
are chosen according to the abscissa of a sufficiently high-order Gauss-Hermite quadrature~\cite{shan-jocs-2016}
i.e., $ \{ ( \omega_i, \mathbf{c}_{i} ) : \, i=1,\, \ldots,\, q \}$, where $\omega_i$ are the quadrature weights. 
It is customary to distinguish between different LBMs using
the D$d$Q$q$ nomenclature, in which $d$ refers to the number of spatial dimensions and $q$ to
the number of discrete components.

The commonly adopted D2Q9 model correctly recovers density and velocity, its 
underlying quadrature is not sufficiently accurate to capture also the temperature \cite{philippi-pre-2006,shan-pre-2010}.
For this reason, in this work, we employ the D2Q17 and the D2Q37 velocity stencils (see Fig.~\ref{fig1:stencils});
while both models can correctly recover the third order velocity moments of the particle distribution function,
it can be shown~\cite{shan-jofm-2006} that the D2Q17 stencils fails to capture the 
non-equilibrium component of the heat-flux $\mathbf{q}$,
\begin{equation}\label{eq:heatflux}
  \mathbf{q} = \frac{1}{2} \sum\limits_{i=1}^q f_i 
                                               \vert  \mathbf{c}_{i} - \mathbf{u} \vert^2 
                                               \left( \mathbf{c}_{i} - \mathbf{u} \right),
\end{equation}
a flaw cured by the D2Q37 stencil.
We consider these two stencils since they consist of the minimal amount of velocities 
allowing to implement seventh- and ninth-order respectively quadrature rules, with non-negative weights 
and integer-valued velocities with a maximum displacement of three~\cite{shan-pre-2010}. 
Since the computational cost of LBM codes generally scales linearly with the number 
of discrete velocities, it is desirable to reduce the velocity set. Pruned version
of these stencils can be obtained by considering only a subset of the moments of the distribution
and/or coupling with regularization procedures~\cite{krivovichev-computation-2023}.

The collision operator $\Omega_i$ is often modeled with the single relaxation time  
Bhatnagar-Gross-Krook (BGK) approximation~\cite{bhatnagar-pr-1954},
\begin{equation}\label{eq:bgk}
  \Omega_i = - \frac{1}{\tau} \left( f_i(\mathbf{x},t) - f_{i}^\eq (\mathbf{x},t) \right),
\end{equation}
consisting of a relaxation with relaxation time $\tau$ towards a local equilibrium $f_{i}^\eq$, which is
defined as expansion in Hermite polynomials of the Maxwell-Boltzmann distribution \cite{shan-jofm-2006}.
In this work, we consider a third order expansion for the D2Q17 stencil,
\begin{flalign}\label{eq:feq-3}
        & f^{\eq,3}_{i}   (\rho, \mathbf{u}, T)   = \,  \omega_i \rho  
        \bigg( 
               1 + \mathbf{u} \cdot \mathbf{c}_{i}  
               & 
               \notag \\
               &\hspace*{2ex} + \frac{1}{2c_s^2} 
               \big[
                    (\mathbf{u} \cdot \mathbf{c}_{i})^2 - u^2 + (T-1) (c_i^2  - 2)
               \big]
               & 
               \notag \\
         & \hspace*{2ex} + \frac{\mathbf{u} \cdot \mathbf{c}_{i}}{6c_s^4} 
               \big[ 
                     (\mathbf{u} \cdot \mathbf{c}_{i})^2 - 3  u^2 + 3(T-1) (c_i^2  - 4)
               \big] 
       \bigg),
\end{flalign} 
and a fourth order expansion for the D2Q37 stencil
\begin{flalign}\label{eq:feq-4} 
        f^{\rm{eq},4}_{i} & (\rho, \mathbf{u},T)   = \, f^{\eq,3}_{i}(\rho, \mathbf{u}, T)   +   \omega_i \rho \bigg(  & \notag \\
            &\frac{1}{24 c_s^6} 
               \bigg[     (\mathbf{u} \cdot \mathbf{c}_{i})^4 
                      - 6 (\mathbf{u} \cdot \mathbf{c}_{i})^2  u^2 + 3  u^4  \notag & \\
        & \quad + 6 (T-1) 
               \big(      (\mathbf{u} \cdot \mathbf{c}_{i})^2 ( c_{i}^2 -  4)
                    + \lvert \mathbf{u} \rvert^2 (4- c_{i}^2) 
               \big)  \notag &  \\
        & \quad + 3 (T-1)^2 ( c_{i}^4 - 8  c_{i}^2 +  8)
               \bigg] 
               \bigg),  
\end{flalign} 
respectively, where  $u^2= \mathbf{u} \cdot \mathbf{u}, 
c_i^2= \mathbf{c}_i \cdot \mathbf{c}_i $ and the speed of sound $c_s$ 
is a lattice specific constant. The values of the lattice weights $\omega_i$ and speed of sound 
for both stencils are given in Appendix B. 

By applying a multiscale Chapman-Enskog expansion, see~\cite{chapman-book-1970}, 
it can be shown~\cite{scagliarini-pof-2010} that Eq.~\eqref{eq:lbe} delivers 
a second order approximation of the macroscopic Navier-Stokes-Fourier equations.
In the absence of external forces, they can be stated as
\begin{flalign}
    & D_t \rho        = - \rho \partial_i u_i &  \notag\\
    & \rho D_t u_i    = - \partial_i P   +  \mu \partial_{jj} u_i + \left( 
                    1 - \frac{1}{c_v} \right) \mu \partial_i \partial_j u_j &  \label{eq:NSF} \\
    & \rho c_v D_t T  = - P \partial_i u_i  + \kappa \partial_{ii} T 
                    + \boldsymbol{\sigma}^{\prime}_{i,j}  \partial_i  u_j,  \notag
\end{flalign}
where $D_t = \partial t + u_i \partial_i $ is the material derivative, 
$\mu$ is the dynamic viscosity, $\kappa$ is the thermal conductivity,  
$\delta_{ij}$ denotes the Kronecker delta and the Einstein summation 
convention is used. 
In two spatial dimensions, the specific heat capacities at constant volume 
and pressure read $c_v=\frac{d}{2}=1$ and $c_p= \frac{d}{2}+1=2$.
The viscous stress tensor is given as 
\begin{equation}
    \boldsymbol{\sigma}^{\prime}_{i,j} =  
    \mu \left(
     \partial_i u_j + \partial_j u_i - \frac{1}{c_v} \delta_{i,j} \partial_k u_k
    \right)
\end{equation}
and the pressure $P$ is linked to density and temperature by an ideal equation 
of state $P = \rho T$. 
The kinematic viscosity $\nu$ and thermal diffusivity $\alpha$ of the 
fluid are related to the relaxation time parameter $\tau$ as \cite{scagliarini-pof-2010}
\begin{equation} 
  \nu = \frac{\mu} {\rho} =  \left( \tau - \frac{1}{2} \right) c_s^2, 
  \ \ 
  \alpha = \frac{\kappa}{\rho c_p} 
         = \left( \tau - \frac{1}{2} \right) c_s^2.
\end{equation}
Using the single relaxation time BGK collision operator \eqref{eq:bgk}, 
the Prandtl number is restricted to $\text{Pr}= \frac{\nu}{\alpha} =1$.

\begin{figure*}[htb]
\vspace{4ex}
\smartdiagramset{
uniform color list=gray for 5 items,
module x sep = 3.5cm,
back arrow disabled=true, text width=2.5cm, module minimum height=2.7cm
}
\smartdiagramadd[flow diagram:horizontal]{     
     Initialize: populations $f_i$,
     Collide: Evaluate right-hand side of Eq.~\eqref{eq:lbe},
     Stream: Assign result to  left hand side of Eq.~\eqref{eq:lbe},
     Apply BC: Set missing post-streaming populations,
     Update macroscopic fields using Eq.~\eqref{eq:macro}
     }{}
\begin{tikzpicture}[overlay]
\draw[additional item arrow type,color=gray] (module5) -- ++(0,2) -| (module2);
\end{tikzpicture}
\caption{Flowchart describing the basic LBM algorithm.
}
\label{fig:flowchart-lbm-algo}
\end{figure*}
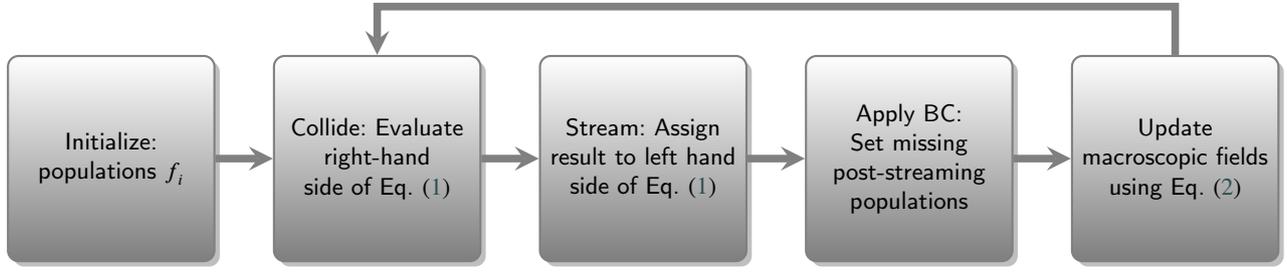

We conclude this section by sketching the LBM algorithm in Fig.~\ref{fig:flowchart-lbm-algo}.
The starting point consists in initializing the discrete distribution $f_i$, for example by 
prescribing initial values for the macroscopic fields via the equilibrium distribution function
(Eq.~\eqref{eq:feq-3} or \eqref{eq:feq-4} in our case).
The LBM iteration consists of alternating the evaluation of the collision step with 
the propagation of the lattice populations along the discrete grid (streaming step) as defined 
by velocity stencil.
Next, missing post-streaming populations at the boundary nodes are prescribed with a suitable BC, 
before updating the macroscopic values to then start the next iteration. 

\section{Characteristic boundary conditions for thermal flows}
\label{sec:cbc}

The general idea of a characteristic BC is to inspect an underlying hyperbolic 
model from a set of macroscopic equations in order to distinguish between 
incoming and outgoing wave components~\cite{thompson-jocp-1987}. 
This basic hyperbolic description is obtained by disregarding viscous 
and tangential boundary terms, giving rise to the locally one dimensional 
inviscid (LODI) approximation~\cite{poinsot-jocp-1992}. 
Next, outgoing wave components resulting from the bulk dynamics are 
left unchanged, while incoming waves are manipulated to achieve a desired behavior. 

In this section, we start by detailing in Sec.~\ref{subsec:background_characteristicBC}
the steps required to define a CBC for the Navier-Stokes-Fourier equations, 
also discussing in Sec.~\ref{subsec:choose_L} a few possible choices for the 
manipulation of the incoming waves.
In Sec.~\ref{subsec:Realization}, we then provide details for the adaptation of the 
CBC to LBM.

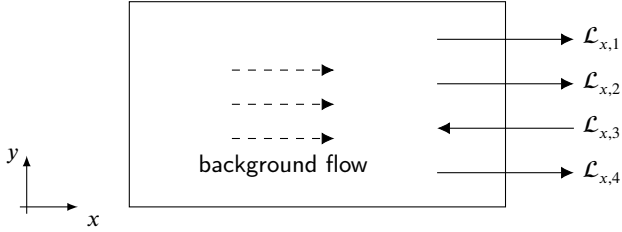
\begin{figure}[htb]          
        \centering
        \begin{tikzpicture}[scale = 0.45]
                         
        \draw (-1, 0) rectangle ( 10., 6);                    
        
        \draw[black,-{Latex[width=1.6mm]}]   (8,1) -- (12,1) node[right] {$\mathcal{L}_{x,4}$};    
        \draw[black,-{Latex[width=1.6mm]}]   (12,2.3) node[right] {$\mathcal{L}_{x,3}$}  -- (8,2.3) ;
        \draw[black,-{Latex[width=1.6mm]}]   (8,3.6) -- (12,3.6) node[right] {$\mathcal{L}_{x,2}$};
        \draw[black,-{Latex[width=1.6mm]}]   (8,4.9) -- (12,4.9) node[right] {$\mathcal{L}_{x,1}$};


        \node at (3.5,1.2)  {background flow};
        \draw[black,dashed,-{Latex[width=1.6mm]}]   (2,4) -- (5,4) ;
        \draw[black,dashed,-{Latex[width=1.6mm]}]   (2,3) -- (5,3) ;
        \draw[black,dashed,-{Latex[width=1.6mm]}]   (2,2) -- (5,2) ;

        \draw[black,-{latex[width=1.5mm]}] (-4.2,0) -- (-2.5,0) node[below right]{$x$};
        \draw[black,-{latex[width=1.5mm]}] (-4,-0.2) -- (-4,1.5) node[left]{$y$};
        \end{tikzpicture}
        \caption{Two dimensional rectangular computational domain, with an outlet 
                 at a right-hand side $x$-boundary. The sketch provides an example 
                 for the orientation of the characteristic waves amplitude 
                 variation $\mathcal{L}_{x,i}$, assuming $u_x>0$ and $\mathrm{Ma} < 1$ .
            }
        \label{fig:orientation-L}
\end{figure}

\subsection{Background on wave amplitudes, LODI and CBC}
\label{subsec:background_characteristicBC}

We assume a bounded rectangular computational domain in $d=2$ dimensions. 
For the sake of brevity, we only discuss the right-hand side boundary 
(i.e., $x=x_b$ and $y$ is inside an interval, cf. Fig.~\ref{fig:orientation-L})
and a procedure for the corners of the computational domain.
However, the treatment of other straight boundaries is straightforward.

Starting from Eqns.~\eqref{eq:NSF}, we can cast the time evolution of
the vector of macroscopic quantities 
$\mathbb{U}:=\begin{pmatrix} \rho, u_x, u_y, T \end{pmatrix}^\top$
as the sum of three distinct contributions
%
%
\begin{equation}\label{eq:NSF-CHAR}
  \frac{\partial \mathbb{U}}{\partial t} 
  = 
  - A \frac{\partial \mathbb{U}}{\partial x} + \mathbb{T} + \mathbb{V}, 
\end{equation}
%
respectively:
\begin{enumerate}
  \item[i)] the term $A \frac{\partial}{\partial x}\mathbb{U}$, which accounts for
        derivatives normal to the boundary,
  \item[ii)] $\mathbb{T}$, which includes spatial derivatives in transversal directions,
  \item[iii)] $\mathbb{V}$, which includes viscous contributions.
\end{enumerate}

The explicit form for these terms reads
\begin{flalign}
	\!\!\!\!\!\!\!
	A = 
	\begin{pmatrix}
		u_x                  & \rho & 0   & 0    \\[0.5ex]
	   \frac{ \tilde{T} }{\rho} & u_x  & 0   & c_s^2 \\[0.5ex]
		0                    & 0    & u_x & 0    \\[0.5ex]
		0                    & \frac{\tilde T}{c_s^2}    & 0   & u_x
	\end{pmatrix}, \\
    \mathbb{T} =  \vek{ -   \frac{\partial (\rho u_y)}{\partial y} \\[0.5ex]
         - u_y \frac{\partial u_x}{\partial y} \\[0.5ex]
         - \frac{1}{\rho} \frac{\partial (\rho \tilde{T})}{\partial y}  - u_y \frac{\partial u_y}{\partial y} \\[0.5ex]
         -  \frac{1}{c_s^2} \frac{\partial (\tilde T u_y)}{\partial y}
     }
\end{flalign}
and 
\begin{flalign}
    &\mathbb{V} =  \vek{ 0 \\[0.5ex]
    \nu \Delta u_x \\[0.5ex]  
    \nu \Delta u_y\\[0.5ex]
    \!
    \frac{\nu c_p}{ \text{Pr} \  c_s^2} \Delta \tilde T + \frac{\nu}{c_s^2} \left( \left( \frac{\partial u_x}{\partial x} - \frac{\partial u_y}{\partial y} \right)^{\!\! 2}
    + \left( \frac{\partial u_x}{\partial y} + \frac{\partial u_y}{\partial x} \right)^{\!\! 2} \right)\!
    }. &&
\end{flalign}
In the above expressions, a rescaled temperature $\tilde{T} \to T c_s^2$ is introduced to ensure 
that the reference temperature is $T_0 = 1$ in lattice units.

A diagonalization of the matrix $A$ results in $A = S^{-1} \Lambda S$  
with 
$\Lambda = \textrm{diag}\left(u_x, u_x, u_x - \sqrt{2 \tilde{T}}, u_x + \sqrt{2 \tilde{T}}\right)$.
The matrices $S$ and $S^{-1}$ are given by

\begin{align*}
    S &= 
    \begin{pmatrix}
    -\frac{\tilde T}{2 \rho c_s^2}  & 0                                   & 0 & \frac{1}{2} \\[0.7ex]
    0                          & 0                                   & 1 & 0           \\[0.7ex]
    \frac{\tilde T}{4 \rho c_s^2}   & -\sqrt{\frac{\tilde T}{8 c_s^4}} & 0 & \frac{1}{4} \\[0.7ex]
    \frac{\tilde T}{4 \rho c_s}   &  \sqrt{\frac{\tilde T}{8 c_s^4}}   & 0 & \frac{1}{4} 
    \end{pmatrix} \quad \text{ and} \\
    S^{-1} &=
    \begin{pmatrix}
    -\frac{\rho c_s^2}{\tilde T} &  0 & \frac{\rho c_s^2}{\tilde T}             & \frac{\rho c_s^2}{\tilde T}   \\[0.7ex]
    0                       &  0 & - \sqrt{\frac{2 c_s^4}{\tilde T}} & \sqrt{\frac{2 c_s^4}{\tilde T}}   \\[0.7ex]
    0                       & 1  & 0                                  & 0 \\[0.7ex]
    1                       & 0  & 1                                  & 1
    \end{pmatrix}. 
\end{align*}
The explicit terms for the three-dimensional case are provided in Appendix A.

With this, the vector of wave amplitude variations for waves crossing the right-hand side boundary is defined as
\begin{equation}
  \mathcal{L}_x= 
      \begin{pmatrix}
         \mathcal{L}_{x,1}, \ \mathcal{L}_{x,2} , \ \mathcal{L}_{x,3}, \ \mathcal{L}_{x,4}
     \end{pmatrix}^\top = \Lambda S \frac{\partial  \mathbb{U}}{\partial x}.
\end{equation} 
 The explicit form of this equation is
%
\begin{equation}\label{eq:thermal_WAV}
    \begin{pmatrix}
    \mathcal{L}_{x,1} \\[0.5ex]
    \mathcal{L}_{x,2} \\[0.5ex]
    \mathcal{L}_{x,3} \\[0.5ex]
    \mathcal{L}_{x,4} 
    \end{pmatrix} 
    =
    \begin{pmatrix}
    u_x   \left( - \frac{\tilde T}{2 \rho c_s^2}   \frac{\partial \rho} {\partial x} + \frac{1}{2 c_s^2}   \frac{\partial \tilde T} {\partial x} \right) \\[0.5ex]
    u_x   \frac{\partial u_y} {\partial x} \\[0.5ex]
    \left( u_x - \sqrt{2 \tilde T}  \right)   \left( \frac{ \tilde T}{4 \rho c_s^2}   \frac{\partial \rho} {\partial x} - \sqrt{\frac{ \tilde T}{8 c_s^4}}   \frac{\partial u_x} {\partial x} +  \frac{1}{4 c_s^2}   \frac{\partial \tilde T} {\partial x} \right) \\[0.5ex]
    \left( u_x +  \sqrt{2 \tilde T}  \right)   \left( \frac{ \tilde T}{4 \rho c_s^2}   \frac{\partial \rho} {\partial x} + \sqrt{\frac{\tilde T}{8 c_s^4}}   \frac{\partial u_x} {\partial x} +  \frac{1}{4 c_s^2}   \frac{\partial \tilde T} {\partial x} \right)
    \end{pmatrix} 
    .
\end{equation} 

From a physical point of view, the eigenvalues
$\Lambda_{1,1}$ and $\Lambda_{2,2}$ represent the speed of the convection wave 
and of advection wave in  $x$-direction respectively \cite{poinsot-jocp-1992}.
$\Lambda_{3,3}$ and $\Lambda_{4,4}$ represent instead the velocities 
of sound waves moving in negative and positive $x$-direction respectively.
Moreover, the orientation of $\mathcal{L}_{x,i}$ is defined by the sign of 
the corresponding eigenvalue $\Lambda_{ii}$, that is, waves propagating along 
(opposite) the $x-$direction correspond to positive (negative) eigenvalues 
(see again Fig.~\ref{fig:orientation-L} for an example).

Now, the outward pointing waves are determined by the bulk dynamics and can thus be computed from 
Eq.~\eqref{eq:thermal_WAV} whereas the inward pointing waves encode information injected into 
the system from outside of the computational domain and need to be specified.
Hence, we need to replace $\mathcal{L}_x$ 
with a vector $\mathcal{\bar{L}}_x$  to modulate inward pointing
wave amplitudes (a few possible choices are discussed in Sec.~\ref{subsec:choose_L}).

Observe that by discarding transversal and viscous terms at the boundary 
Eq.~\eqref{eq:NSF-CHAR} reduces to
\begin{equation}\label{eq:thermal_LODI_macro_evo}
  \frac{ \partial \mathbb{U}}{ \partial t} = -S^{-1} \mathcal{\bar{L}}_x ,
\end{equation}
which coincides with the LODI approximation.

The CBC approach~\cite{yoo-ctan-2007,jung-jocp-2015}, instead, aims at 
including  the effect of 
transversal and viscous contributions at the boundary to the time evolution of  $\mathbb{U}$ by solving
\begin{equation}\label{eq:thermal_CBC_macro_evo}
    \frac{ \partial \mathbb{U}}{ \partial t} = 
      -S^{-1}  \mathcal{\bar{L}}_x + \mathbb{T} + \mathbb{V}.  
\end{equation}

\subsection{Choices for incoming wave amplitudes}
\label{subsec:choose_L}
%
In this section, we revise possible strategies for the treatment of 
incoming wave amplitudes for a CBC.
As discussed in the previous section, the orientation of waves crossing the boundary is given 
by the sign of the corresponding eigenvalue $\Lambda_{ii}$. 
Therefore, external knowledge about the sign of $u_x$ is required to obtain the time evolution of $\mathbb{U}$ (and therefore $u_x$).
In a discretized scheme (see Sec.~\ref{subsec:Realization}), the sign of $u_x$ at time $t+\Delta t$ is taken as the sign at time $t$.

\noindent\paragraph{A) Annihilation.\/} 
A common approach consists of choosing incoming wave amplitudes such that their
contribution to the time evolution of $\mathbb{U}$ 
vanishes~\cite{hedstrom-jocp-1979,thompson-jocp-1987}. In other words,
this means no information enters the bulk and the influence of external dynamics 
on the domain of interest is completely suppressed.

In the LODI approximation (Eq.~\eqref{eq:thermal_LODI_macro_evo}), 
this translates to setting incoming wave amplitude variations to zero, i.e., 
we substitute $\mathcal{L}_x$ with a vector $\mathcal{\bar{L}}_{x}$, 
whose $i-$th component is defined as
\begin{equation}\label{eq:PNR}
	\mathcal{\bar{L}}_{x,i} = \begin{cases}
		                          \mathcal{L}_{x,i} \;\text{ for an outgoing wave $i$,} \\
		                                 0 \quad \  \;\text{ for an incoming wave $i$}.
	                          \end{cases}
\end{equation}

By contrast, in the CBC approach, setting incoming wave amplitudes to zero will not 
guarantee that no information will travel from the boundary to the bulk domain
for cases where transversal and viscous contributions are relevant to the dynamic.
This can be seen by casting Eq.~\eqref{eq:thermal_CBC_macro_evo} in the following form:
\begin{equation}
	\frac{ \partial \mathbb{U}}{ \partial t} 
   = -S^{-1}       \mathcal{\bar{L}}_x + \mathbb{T} + \mathbb{V}
	 = -S^{-1} \big( \mathcal{\bar{L}}_x - \mathcal{T}_x - \mathcal{V}_x \big),
\end{equation}
where $\mathcal{T}_x = S \mathbb{T}, \quad \mathcal{V}_x = S \mathbb{V}$.

As a remedy, the contributions $\mathcal{T}_x$ and $\mathcal{V}_x$ 
can be absorbed in the unknown incoming wave amplitude variation $\bar{\mathcal{L}}_{x,i}$ 
as proposed in Ref.~\cite{jung-jocp-2015}: 
\begin{equation}\label{eq:CBC-PNR}
	\bar{\mathcal{L}}_{x,i} = \mathcal{T}_{x,i} + \mathcal{V}_{x,i}.
\end{equation}

This strategy of completely annihilating incoming waves theoretically leads to
a perfectly non-reflecting BC. In practice, however,
due to discretization errors and the fact that wave 
amplitudes get computed from an approximate system, reflection waves
are generally still present.

\noindent\paragraph{B) Relaxation toward target quantities.\/}
As observed in the previous paragraph, posing a perfectly non-reflecting BC 
gives no control over the macroscopic values at the boundary since their 
time evolution strongly depends on the outgoing waves. 
On the other hand, imposing desired target values by means of a Dirichlet BC 
generally leads to significant reflection waves.
As a trade-off between these two cases, a relaxation towards a target
macroscopic value can be incorporated in the incoming wave amplitude variations~\cite{rudy-jocp-1980,poinsot-jocp-1992}.
A general expression for the unknown wave amplitudes in conjunction with the CBC 
approach was proposed in Ref.~\cite{yoo-ctan-2007} and reads as:
\begin{equation}\label{eq:CBC-Relax}
	\mathcal{\bar{L}}_{x,i} = \mathcal{T}_{x,i}  
                          + \mathcal{V}_{x,i} 
                          + \alpha (\mathcal{T}_{x,i}^{\infty} - \mathcal{T}_{x,i})
                          + \beta ( Z - Z_{\infty}),
\end{equation}
where a chosen macroscopic quantity $Z$ (e.g. the pressure), and transversal 
waves $\mathcal{T}_{x,i}$, are relaxed towards 
target values $Z_{\infty}$ and $\mathcal{T}_{x,i}^{\infty}$ at rates 
$\alpha$ and $\beta$, respectively.
This strategy has been reported to increase numerical stability and accuracy \cite{yoo-ctan-2007}.
However, the relaxation coefficients pose additional degrees of freedom that have to be determined.

Let us remark that the same strategy can be applied to the LODI approach 
\eqref{eq:thermal_LODI_macro_evo}, i.e., 
\begin{equation}\label{eq:LODI-Relax}
    \mathcal{\bar{L}}_{x,i} = \alpha (\mathcal{T}_{x,i}^{\infty} - \mathcal{T}_{x,i}) 
                            + \beta ( Z - Z_{\infty}).
\end{equation}

\subsection{Realization of characteristic BC in the LBM} 
\label{subsec:Realization}
%
In the previous sections, we described how to obtain  target values at the macroscopic level.
We now describe how to pose these target values in a multi-speed LBM to implement a characteristic BC.

\begin{figure*}[htb]
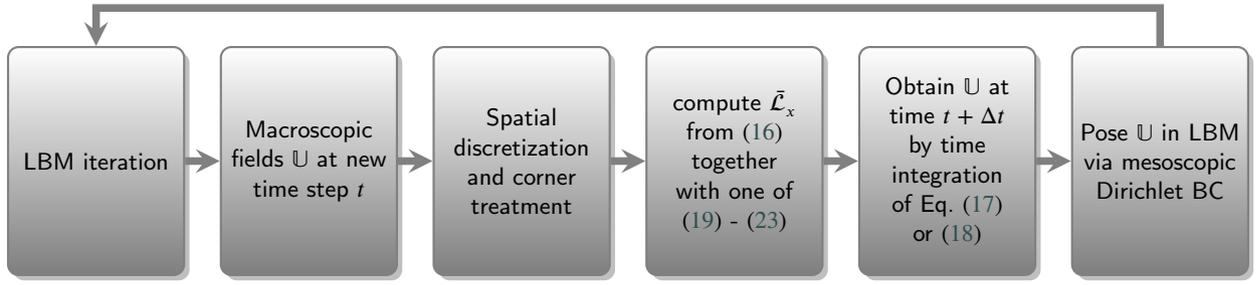


\smartdiagramset{
uniform color list=gray for 6 items,
module x sep = 2.8cm,
back arrow disabled=false, text width=2.1cm, module minimum height=3.cm
}
\smartdiagram[flow diagram:horizontal]{
    LBM iteration,
    Macroscopic fields $\mathbb{U}$ at new time step $t$, 
    Spatial discretization and corner treatment, 
    compute $\mathcal{\bar{L}}_{x}$ from \eqref{eq:thermal_WAV} 
    together with one of  \eqref{eq:PNR}\ -  \eqref{eq:LODI-Relax}, 
    Obtain $\mathbb{U}$ at time $t+\Delta t$ by time integration of 
    Eq.~\eqref{eq:thermal_LODI_macro_evo} or \eqref{eq:thermal_CBC_macro_evo}, 
    Pose $\mathbb{U}$ in LBM via mesoscopic Dirichlet BC
  }
\caption{Flowchart of the conceptual steps required to pose a CBC in the LBM. }
\label{fig:flowchart-cbc-lbm}
\end{figure*}

The general procedure~\cite{izquierdo-aps-2008,heubes-jcam-2014,jung-jocp-2015} 
is summarized in Fig.~\ref{fig:flowchart-cbc-lbm}.
The starting points are the macroscopic flow fields computed by the LBM algorithm
(see Sec.~\ref{sec:lbm}) at a generic time $t$. The task of the boundary condition
is to define the lattice populations left undefined at the boundary of the 
computational domain. To this aim, we perform a spatial discretization, replacing
the spatial derivatives with finite differences and enabling the 
computation of the discrete analog of the vector of manipulated wave amplitudes  
$\mathcal{\bar{L}}_x$ given by Eq.~\eqref{eq:thermal_WAV}. 
Next, we plug this vector into the corresponding macroscopic evolution  
and perform a time integration, delivering the macroscopic target values 
for the next time step $t+\Delta t$. 
Finally, the computed target values are supplied to the LBM by means of 
a mesoscopic Dirichlet BC, thus specifying the missing populations at time $t+\Delta t$.
%

In the remaining part of this section, we provide details on 
the implementation of CBC for multi-speed stencils, corner treatment
and possible choices for space and time discretization.

\begin{figure}[htb]
\begin{center}
\begin{tikzpicture}[scale=0.7]

    \draw [dashed]   (3.5, 0.5) rectangle (6.5,3.5);      

    \foreach \y in {1,...,3}{
        \foreach \x in {1,2,3,4,5,6}
        {
            \draw[fill] (\x, \y) circle(3pt) ;
        }
    }
    \foreach \y in {-2,-1,0}{
        \foreach \x in {4,5,6}
        {
            \draw[fill] (\x, \y) circle(3pt) ;
        }
    }

    \foreach \y in {-2,-1,0}{
        \foreach \x in {1,2,3}
        {
            \draw (\x, \y) circle(3pt) ;
        }
    }        
    \node[mark size=3.pt] at (4,1) {\pgfuseplotmark{square*}};
    
    \draw[-{Latex[width=1.6mm]}, thick] (3.4,0.4) -- (2.3,-0.7) node[midway, below]{$v$} ;

          \node (A) at ( 3,-3) [scale=1.1] {$x_{f}$};
          \node (B) at ( 4,-3) [scale=1.1] {$x_{b,1}$};
          \node (C) at ( 5,-3) [scale=1.1] {$x_{b,2}$};
          \node (D) at ( 6,-3) [scale=1.1] {$x_{b,3}$};

\end{tikzpicture}
\end{center}
\caption{Schematic boundary geometry for multi-speed velocity stencils with a 
         displacement of $M=3$. Filled (hollow) symbols denote boundary (bulk) nodes. 
         To pose a characteristic BC, finite differences are applied to approximate 
         spatial derivatives in the boundary nodes. 
         The square node is used to calculate the target macroscopic quantities 
         for all corner nodes in the dashed rectangle. 
         In this case, spatial derivatives are evaluated along the 
         inward diagonal $v$ indicated by the arrow. 
         }
\label{fig2:Layer-Multispeed-BC-v2}
\end{figure}
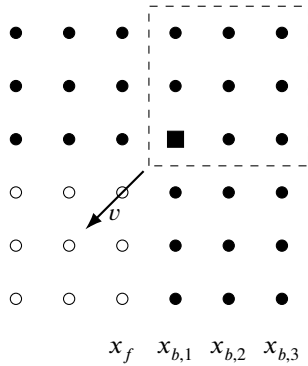

\paragraph{Multi-speed LBM.\/}
\noindent For multi-speed LBM, $M$ layers of boundary nodes exhibit missing 
populations, where $M$ is the maximum displacement of the underlying velocity stencil. 
In this work, we consider the D2Q17 and D2Q37 velocity stencils (Fig.~\ref{fig1:stencils}), 
both having maximum displacement $M=3$. 
Let us label the boundary nodes at the right-hand side boundary, 
for a fixed $y$, as $\mathbf{x}_{b,j}=(x_{b,j},y), \ j~=1, \ldots, M$, 
where $\mathbf{x}_{b,1}$ is adjacent to the rightmost bulk node 
$\mathbf{x}_{f}=(x_{f} ,y)$ and $\mathbf{x}_{b,M}$ is the outermost boundary node   
(see Fig.~\ref{fig2:Layer-Multispeed-BC-v2}).
The characteristic analysis is conducted for each layer of boundary nodes. 
As explained in Sec.~\ref{subsec:background_characteristicBC}, 
waves crossing the $j-$th layer of boundary 
nodes are identified by the sign of the corresponding eigenvalue and  
the incoming wave amplitude variations are posed on $\mathbf{x}_{b,j}$. 
Note that the resulting target values may differ for the various layers forming the boundary.
Finally, the macroscopic equation describing the time evolution of $\mathbb{U}$ 
on the boundary is solved numerically.
Details on the numerical solution and the enforcement of $\mathbb{U}$ on the boundary nodes are reported in the paragraphs below.
%

\paragraph{Spatial discretization and corner treatment.\/}
The spatial derivatives of macroscopic quantities $\mathbb{U}$ on a boundary 
node $\mathbf{x}_{b,j}$ at a fixed time $t$ are approximated with second order finite differences. 
Dropping the fixed time $t$ for the sake of a compact notation, 
we denote a spatial discretization step in $x-$direction by  
$\mathbf{e}_x = (\Delta x, 0)^\top$. For $\mathbf{x}_{b,M}$ (outermost layer), 
we use  one-sided differences for the spatial derivatives:
\begin{flalign}
     & \frac{\partial \mathbb{U}_i(\mathbf{x}_{b,M})} {\partial x} 
     \approx
      \tfrac{ 1}{2} \bigl(  
            3 \mathbb{U}_i( \mathbf{x}_{b,M}) 
            \!-\! 4 \mathbb{U}_i(\mathbf{x}_{b,M-1}) 
            +  \mathbb{U}_i(\mathbf{x}_{b,M-2})    
    \bigr) \label{eq:FD-dx-bw}.\!\!\!  &&
\end{flalign}
For inner boundary nodes $\mathbf{x}_{b,j}, \ j =1,2,\dotsc, M-1$,  
we use central finite differences for the derivatives: 
\begin{flalign}
     \frac{\partial \mathbb{U}_i(\mathbf{x}_{b,j})} {\partial x} & \approx 
      \tfrac{ 1}{2} \bigl(  
               \mathbb{U}_i(\mathbf{x}_{b,j} +  \mathbf{e}_x)
            -  \mathbb{U}_i(\mathbf{x}_{b,j} -  \mathbf{e}_x)
    \bigr). \label{eq:FD-dx-center} &&
\end{flalign}  
Spatial derivatives in the $y-$direction are evaluated analogously.

We use this procedure for any straight boundary (top/bottom left/right BC, c.f. Fig.~\ref{fig:orientation-L}).
In principle, also corners can be treated using a combination of the LODI 
approaches along the $x-$ and $y-$ axis respectively.
However, this in general requires the definition of compatibility conditions
for cases where the corner lies at the intersection between different 
types of BC~\cite{lodato-jocp-2008}.
In this work we evaluate instead a simpler approach, where 
spatial derivatives are computed in the direction of the inward facing diagonal
(this would be direction $v=(-1,-1)^\top$ in the example given in Fig.~\ref{fig2:Layer-Multispeed-BC-v2},
where we consider the top right corner).
We consider a coarse grained approach, where target macroscopic quantities 
for all corner sites (dashed box in Fig.~\ref{fig2:Layer-Multispeed-BC-v2}) 
are obtained by conducting the characteristic analysis only in 
innermost corner node (square node in Fig.~\ref{fig2:Layer-Multispeed-BC-v2}).

\paragraph{Evaluation of viscous terms.\/}
Using the CBC approach described in this work (see Tab.~\ref{tab:CBCs}), 
we aim to reconstruct the Navier-Stokes-Fourier equations on the boundary.
This should be contrasted with the recent implementation of characteristic 
BCs for multi-speed LBM given in \cite{chen-jocp-2023}, 
where the focus was on acoustic problems and thus the viscous terms were discarded.

To ensure consistency in the coupling of mesoscopic and macroscopic scales at the boundary, 
we  make use of the link between the scales provided by the Chapman-Enskog expansion. 
This multiscale expansion offers expressions for the viscous terms on the macroscopic 
scale in terms of the mesoscopic distribution.

That is, the Laplacian of velocity appearing in Eq.~\eqref{eq:NSF} is  approximated as \cite{kruger-book-2017}
\begin{equation}\label{eq:LaplaceU}
    \mu \partial_{jj} u_k  \approx  \nabla_j \cdot \left(   - \left(1 -\tfrac{1}{2\tau}\right)\sum\limits_{i=1}^{q} c_{i,j} c_{i,k} f_{i}^{\rm{neq}} \right)\!\! ~,
\end{equation}
where the derivatives of the  non-equilibrium part $f_{i}^{\rm{neq}}= f_{i}^{\rm{eq}} - f_{i}$ 
are evaluated in $\mathbf{x}_f$ -- i.e. in the bulk node adjacent to the boundary 
(see Fig.~\ref{fig2:Layer-Multispeed-BC-v2}) -- using the finite differences \eqref{eq:FD-dx-bw} 
and \eqref{eq:FD-dx-center} (along y).
Furthermore, making use of Eq.~\eqref{eq:heatflux}, the Laplacian of the temperature is restated as 
\begin{equation}\label{eq:LaplaceT}
-\kappa \partial_{jj} T = \mathrm{div}(\mathbf{q}) 
                        = \nabla \cdot \frac{1}{2} \sum\limits_{i=1}^q f_i 
                                                                     \vert  \mathbf{c}_{i} - \mathbf{u} \vert^2 
                                                                     \left( \mathbf{c}_{i} - \mathbf{u} \right) .
\end{equation}
For a sufficiently high order quadrature, which allows recovery of the third order moment 
of the distribution, it is then possible to compute the above quantity and to then approximate
the first order spatial derivatives of $\mathbf{q}$ using finite differences 
(e.g. \eqref{eq:FD-dx-bw} or \eqref{eq:FD-dx-center}).

To conclude, in Table~\ref{tab:CBCs} we summarize the different CBC schemes which will be evaluated 
in numerical simulations in the upcoming sections.
\begin{table}[h]
    \begin{tabular}{cccc}
        Name & Macroscopic Eq. & Incoming amplitude  \\
        \hline
        LODI & \eqref{eq:thermal_LODI_macro_evo} & \eqref{eq:PNR} \\
        LODI-RELAX &\eqref{eq:thermal_LODI_macro_evo} & \eqref{eq:LODI-Relax} \\
        CBC & \eqref{eq:thermal_CBC_macro_evo} & \eqref{eq:CBC-PNR} \\
        CBC-RELAX  & \eqref{eq:thermal_CBC_macro_evo} & \eqref{eq:CBC-Relax} \\
    \end{tabular} 
    \caption{Summary of the characteristic based BC considered in this work. }
    \label{tab:CBCs}
\end{table}

\paragraph{Time integration.\/}
The implementation of a characteristic BC requires time integration of 
either Eq.~\eqref{eq:thermal_LODI_macro_evo} or Eq.~\eqref{eq:thermal_CBC_macro_evo}.
As pointed out in Ref.~\cite{chen-jocp-2023}, a simple explicit Euler scheme is not a viable 
option in this case since the coupling with LBM leads to the violation of the CFL condition
for the FD solver.
Therefore, we make use of a fourth order Runge-Kutta scheme (RK4) \cite{stoer-book2002}, 
which requires derivative information at time $t+ \frac{\Delta t}{2}$ approximated using the 
second order finite differences shown above. 
Macroscopic quantities located on bulk nodes at this intermediate stage are obtained 
by linear interpolation in time.

For simplicity, for time integration of the CBC scheme we fix the viscous 
terms \eqref{eq:LaplaceU} and \eqref{eq:LaplaceT} to their values calculated at time $t$, 
since no information about $f$ at intermediate time steps $t+\frac{\Delta t}{2}$ is known. 

\paragraph{Mesoscopic BC.\/}
After the target values $\mathbb{U}$ for time $t+\Delta t$ have been obtained, 
they are enforced in the LBM by means of a Dirichlet BC. 
In this work, two simple ways to do this are considered: i) the equilibrium BC, 
where all populations on $\mathbf{x}_{b,j}$  are set according to the discrete 
expansion of the equilibrium distribution chosen, e.g. according to Eq.~\eqref{eq:feq-3} 
or Eq.~\eqref{eq:feq-4} and ii) the constant non-equilibrium extrapolation BC (NEEP), 
where the non-equilibrium part  of the  bulk node $\mathbf{x}_f$ adjacent to 
the boundary is added to the equilibrium computed on the boundary nodes. 
That is, populations on the boundary node $\mathbf{x}_{b,j}$ are computed as
(see Fig.~\ref{fig2:Layer-Multispeed-BC-v2} for notation  $\mathbf{x}_{f}$ and $\mathbf{x}_{b,j}$)
\begin{flalign}\label{eq:NEEP}
  f(\mathbf{x}_{b,j}, &t+\Delta t) 
  &&\notag
  \\
  &= 
  f^{\eq}(\mathbf{x}_{b,j},t+\Delta t)) 
  + 
  f^{\mathrm{neq}}(\mathbf{x}_{f},t+\Delta t)) ,
  &&
\end{flalign}
where $j=1,\ldots,M$ and 
\begin{equation}
	\!
    f^{\mathrm{neq}}(\mathbf{x}_{f},t+\Delta t))  = 
           f(\mathbf{x}_{f},t+\Delta t) 
           - 
           f^{\eq}(\mathbf{x}_{f},t+\Delta t). 
\end{equation}

\section{Numerical Results}\label{sec:numerics}

In this section, we benchmark accuracy and stability of the characteristic 
boundary condition described in the previous section.
We consider three different numerical experiments. 
In Sec.~\ref{subsec:1dstep}, we take into consideration the one-dimensional 
dynamics of shock waves originated by a smoothed temperature step. 
In Sec.~\ref{subsec:vortex}, we consider the propagation of a vortex out of the computational domain. 
In this benchmark, transversal information becomes relevant also at the boundary and thus, 
there is a significant deviation from the locally one-dimensional assumption used to calculate 
the outgoing wave amplitude variations in the LODI approximation. 
Finally, in Sec.~\ref{subsec:ang-wave}, we inspect the interaction of an planar oblique 
wave with the boundary at various angles and measure the reflection. 
In this setup, the importance of transversal terms can be controlled 
by the initial angle between the wave front and the $y-$axis.

For all the cases above, we compare the performance of different LODI and CBC
realizations against the results provided by a simple zero gradient BC (ZG), 
where populations at the boundary nodes $\mathbf{x}_{b,j}, \  j= 1, \dotsc, M$ are set with 
the values from the nearest the bulk node $\mathbf{x}_f$: 
\begin{equation}\label{eq:zg-bc}
    f_i(\mathbf{x}_{b,j},t+\Delta t) = f_i(\mathbf{x}_f,t+\Delta t),   \quad i =1,\; \ldots,\; q.
\end{equation}

The accuracy of the BCs is quantified as follows: 
For the first two benchmark problems, reference fields $Z^{\text{ref}}$ are obtained 
from a fully periodic LBM simulation on an extended grid for  macroscopic variables $Z \in \{ \rho, u_x, T \}$. 
The extended grid has been chosen sufficiently large, such that no interaction takes place 
between the boundaries and the bulk dynamics in the region of interest.
We then compute i) global relative $L^2$-errors $e_Z$ 
and ii) pointwise relative errors $\widetilde{e}_Z$ with respect to the  reference fields.
They are defined as
\begin{flalign}
    &e_Z = \left( \sum_{(x,y)\in L_x\times L_y} \left( \frac{| Z(x,y) - Z^{\myref}(x,y)|}{|Z^{\myref}(x,y)|} \right)^{\!\! 2} \right)^{\frac{1}{2}}\!, \\
    &\widetilde{e}_Z(x,y) = \frac{|Z(x,y) - Z^\myref(x,y)|}{ |Z^\myref(x,y)|}.
    \label{eq:errors}
\end{flalign}

\subsection{1D temperature step}\label{subsec:1dstep}
\begin{figure}[h]
    \centering
    \includegraphics[width=\linewidth]{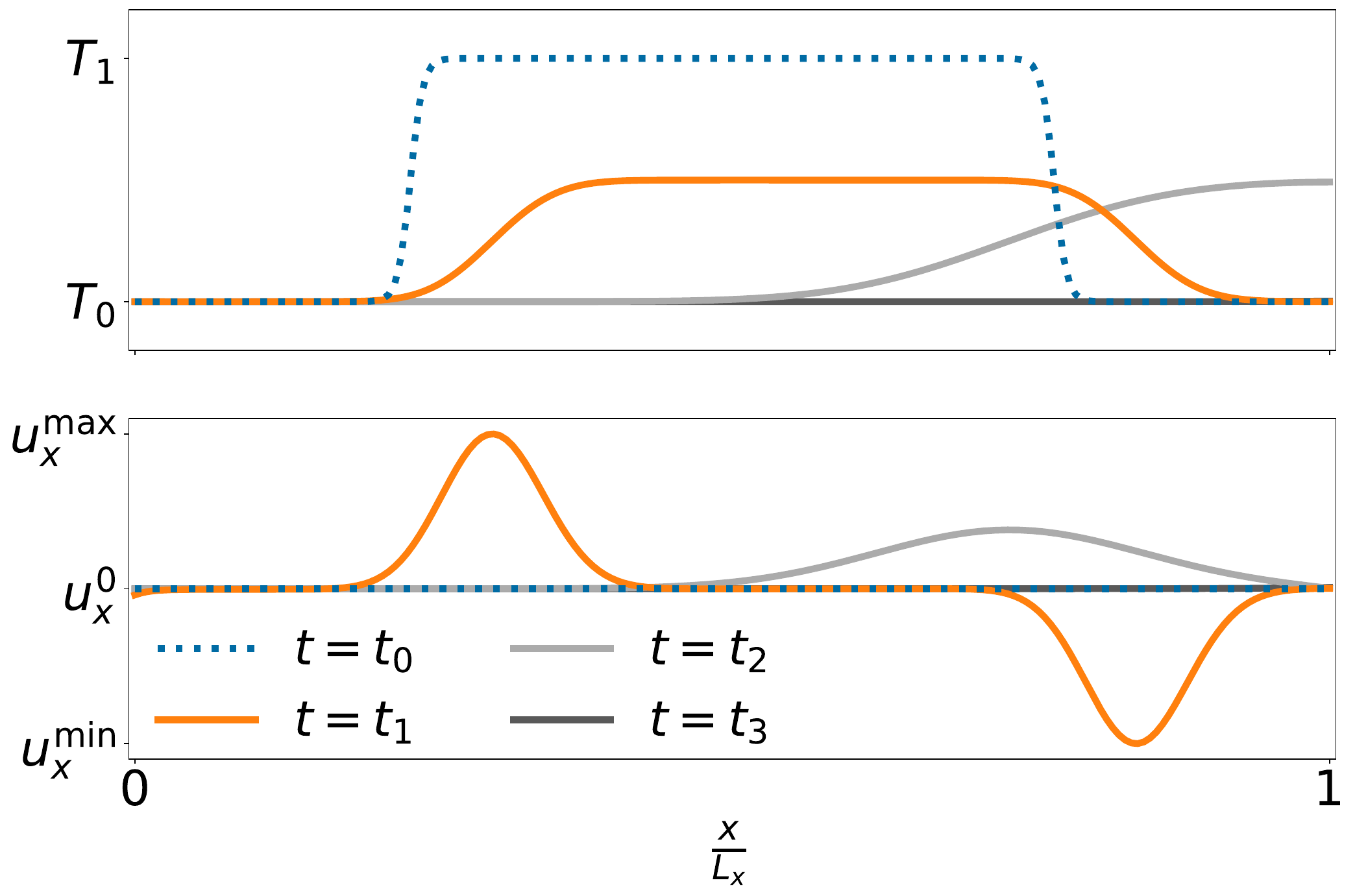}
    \caption{ Temperature and stream-wise velocity profiles of the reference simulation at 
              selected time steps, plotted along the horizontal centerline $y=\frac{L_x}{2}$. 
              The initial configuration at $t=t_0$ is obtained with the following parameters: 
              $\rho_0 = T_0 = 1,  T_1 = 1.0005, \ \mathbf{u_0}=\vek{\text{Ma} \cdot c_s,&0}^\top, \mathrm{Ma} = 0.1, \text{s}=0.5$.  
              The times $t_1$ and $t_2$ correspond to iterations shortly before and during the  
              pulses interaction with the artificial boundary  posed at  
              $x=L_x$ in the simulation on the truncated grid. 
              At $t=t_3$, the system is almost completely at rest (see also Fig.~\ref{fig:thermal-step-L2} for the localization of the time instances $t_i$).
            }\label{fig:thermal-step-ref-temperature}
\end{figure}

We consider a 2D bounded rectangular domain with an initial homogeneous 
density $\rho_0$, a homogeneous background velocity $\mathbf{u}_0 = \vek{\text{Ma} \cdot c_s,&0}^\top$ (parallel to the $x$-axis) at $\mathrm{Ma}=0.1$ and a smooth temperature step (parallel to the $x$-axis, 
for temperature $T_0$ to $T_1$ and back to $T_0$ ) due to a hyperbolic tangent, where the steepness parameter $s$ controls the slope of the step. Formally:
\begin{equation}
    \rho(x,y) = \rho_0, \quad  \mathbf{u}(x, y) = \mathbf{u_0}
\end{equation}
and
\begin{equation}
    T(x,y) = \begin{cases}
                    T_1 + \frac{T_1-T_0}{2}  ( \tanh \left( \text{s} \left(x-\tfrac{L_x}{4}\right)  \right) -1), \text{ if } x \leq \frac{L_x}{2}  \\
                    T_1 - \frac{T_1-T_0}{2}  ( \tanh \left( \text{s} \left(x-\tfrac{3L_x}{4}\right)  \right) -1), \text{ else}.
                \end{cases} 
\end{equation}
%
The initial conditions and the corresponding dynamic is sketched 
in Fig.~\ref{fig:thermal-step-ref-temperature}. 
The specific numerical values are found in the caption of Fig.~\ref{fig:thermal-step-ref-temperature}.

The left- and right-hand side boundaries of the rectangular domain are equipped with artificial BC, 
while the upper and lower boundaries are taken to be periodic, 
thus we ignore here error contributions coming from the treatment of corner nodes.

For this flow, the LODI approximation underlying the computation of the wave 
amplitude variations is well justified: the flow is globally one dimensional 
and only viscous terms are discarded in the approximation.

Numerical simulations have been conducted using the D2Q17 stencil 
on a grid of size $L_x \times L_y = 200\times 20$ and the relaxation 
time used in simulation was $\tau=0.9$ in numerical units.
Fig.~\ref{fig:thermal-step-ref-temperature} depicts the reference simulation 
(for $T^\myref$ and $u_x^\myref$) in the region of interest at selected times 
(from $t_0$ initial value till $t_3$, where the system almost reaches a resting state).

\begin{figure}[h]
  \centering
  \includegraphics[width=\linewidth]{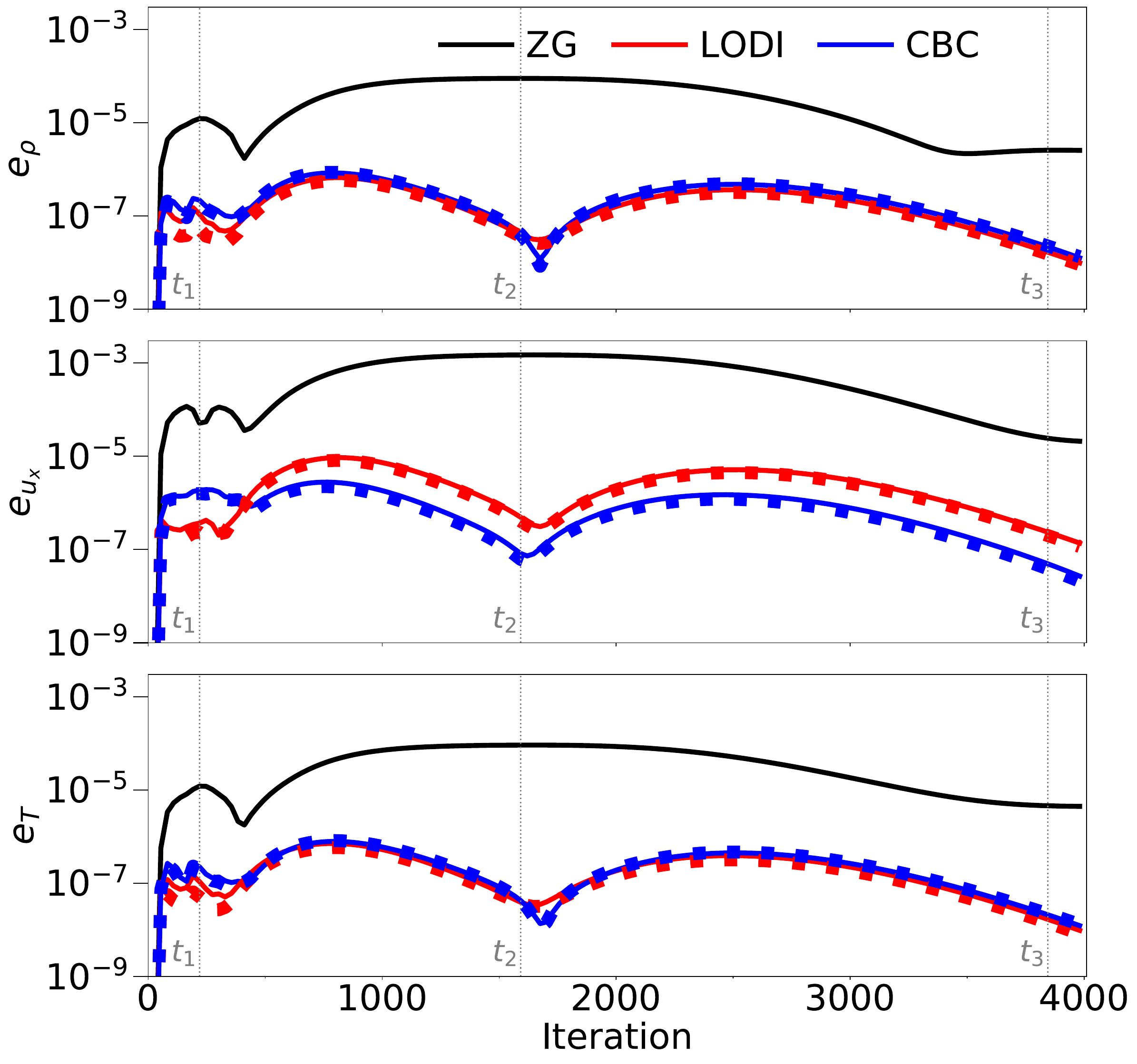}
  \caption{Evolution of $e_Z$ in the thermal step benchmark with $\tau=0.9$. 
            The LODI and CBC schemes are applied with the equilibrium BC \eqref{eq:feq-3} (solid lines) and NEEP BC \eqref{eq:NEEP} (dashed lines).   
           The vertical dotted lines correspond to the times 
           $t_1, t_2$ and $t_3$ in Fig.~\ref{fig:thermal-step-ref-temperature}. 
           }\label{fig:thermal-step-L2}
\end{figure}

In Fig.~\ref{fig:thermal-step-L2}, we present the time evolution of the global relative $L^2$-errors $e_Z$.
Evidently, any characteristic BC reduces the value of $e_Z$ for all macroscopic quantities 
taken into consideration by two or more orders of magnitude on average. 
Since transversal terms are negligible for this benchmark, the only difference between the LODI 
and CBC implementations is the incorporation of viscous terms. 
It is observed that the  CBC scheme leads to slightly larger errors for the first $400$ iterations 
(i.e., before significant interaction with the boundary sets in), see Fig.~\ref{fig:thermal-step-L2}. 
As explained in Sec.~\ref{subsec:Realization}, this difference comes from the fact that 
in the RK4-scheme used for time integration of Eq.~\eqref{eq:thermal_CBC_macro_evo}, 
no value for the heat-flux is available at the intermediate time $t+ \frac{\Delta t}{2}$ 
and the old value from time $t$ is used instead.
As the impinging wave starts interacting with the boundary, the error levels obtained with the 
CBC  coincide with those of the  LODI schemes in $e_{\rho}$ and $e_{T}$ and improve in $e_{u_x}$.
In Fig.~\ref{fig:thermal-step-ref-temperature}, we also compare the results obtained employing 
different types of Dirichlet BC for imposing the macroscopic values. For this numerical setup,
no major differences can be observed comparing the equilibrium BC and the NEEP BC.

\begin{figure}[htb]
    \centering
    \includegraphics[width=\linewidth]{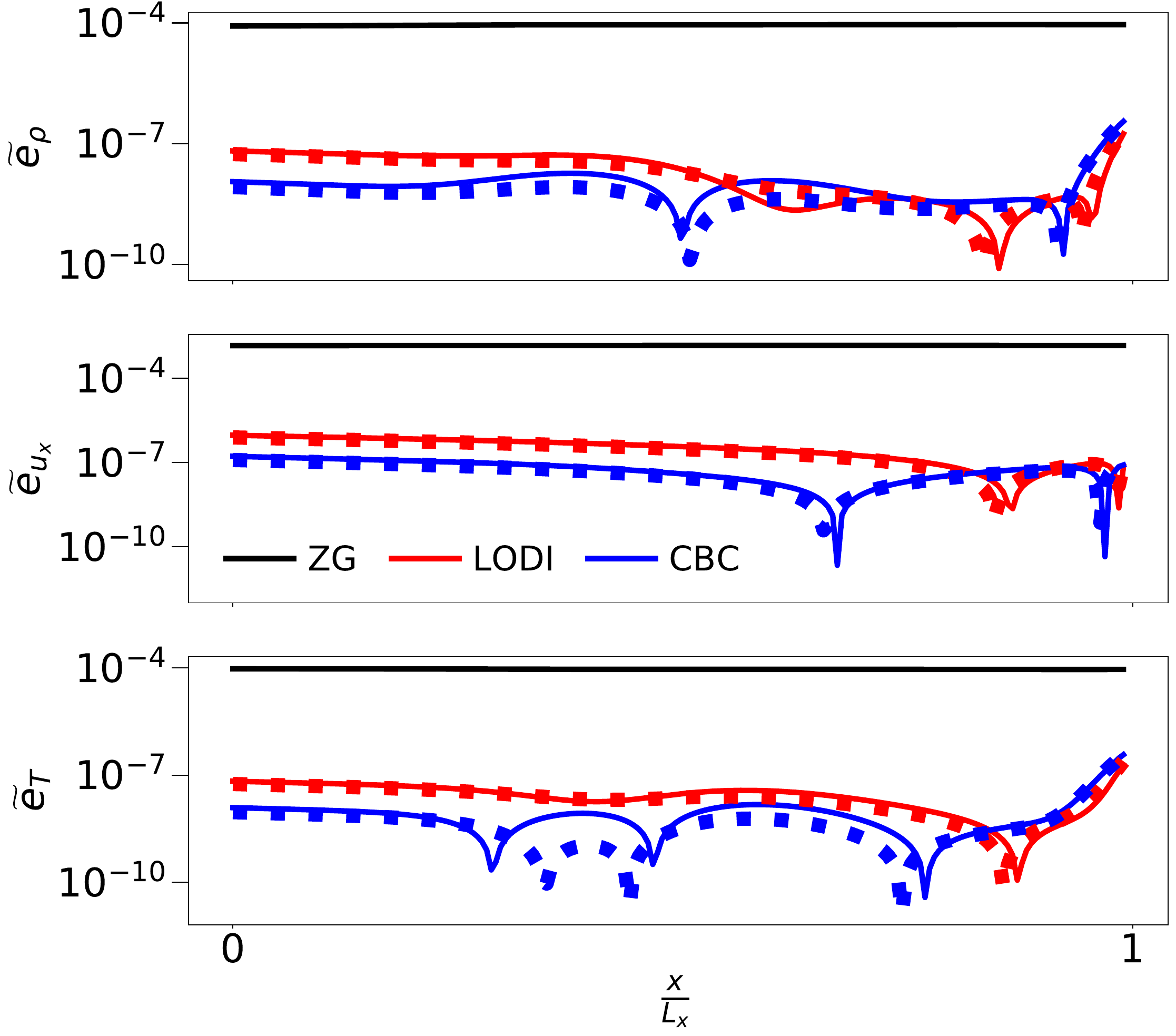}
    \caption{Pointwise relative errors $\widetilde{e}_Z$ at time $t_2$ along the slice $y=\frac{L_y}{2}$.
    The LODI and CBC schemes are applied with the equilibrium BC \eqref{eq:feq-3} (solid lines) and NEEP BC \eqref{eq:NEEP} (dashed lines).   
    }
    \label{fig:thermal-step-rel-err-pointwise2}
\end{figure}

In  Fig.~\ref{fig:thermal-step-rel-err-pointwise2}, the pointwise relative errors $\widetilde{e}_Z$  
along the horizontal midplane $y=\frac{L_y}{2}$ are shown at time $t_2$.
Consistent to the global $L^2$-errors, the ZG produces pointwise errors almost three 
orders of magnitude higher in all the points considered when compared to the characteristic schemes. 
In particular, the usage of the ZG scheme has a larger impact on the bulk dynamics than any 
characteristic based BC; we note that also the accuracy in the center of the computational domain, 
away from inlet and outlet, is degraded.
The LODI and CBC schemes yield very similar pointwise errors that are maximized close 
to the outlet, where small oscillations are introduced. 
Again, the differences between NEEP and equilibrium BCs are negligible.

\subsection{Propagating Vortex}\label{subsec:vortex}
\begin{figure}[htb]
    \centering
    \includegraphics[width=\linewidth]{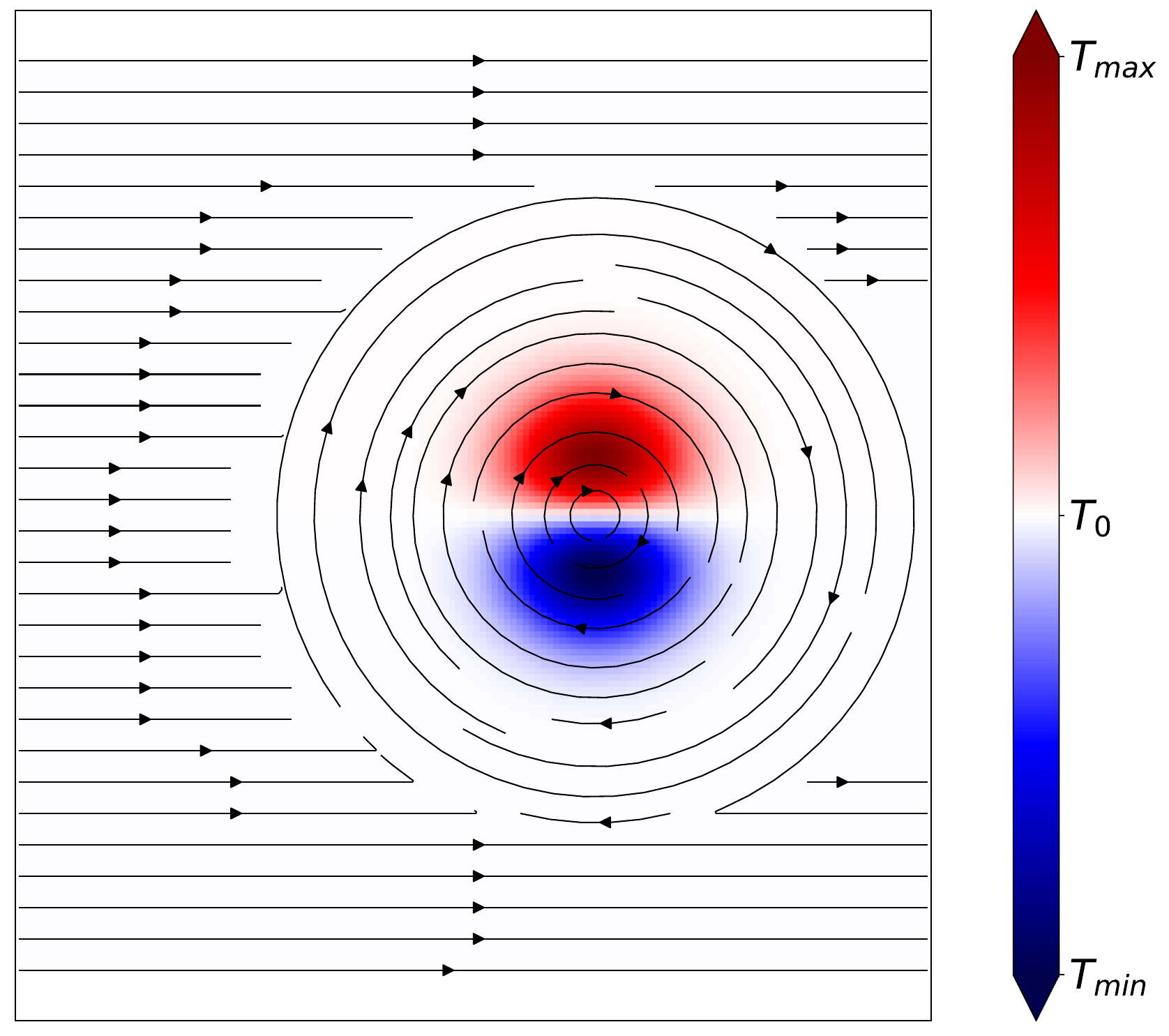}
    \caption{Initial setup for thermal vortex benchmark in the computational domain. 
             Temperature $T(x,y)$ (heat map) varies around $T_0 = 1$ in a circle with center 
             $(\hat{x}_0,\hat{y}_0)$ with radius $\hat{r}=0.7$ and strength $b=\frac{3}{20}$, 
             see \eqref{eq:initialvalues-vortex-example}. 
             The initial velocity field (streamlines) is the superposition with 
             a global background velocity $\mathbf{u_0} = \vek{\text{\normalfont Ma} \cdot c_s,&0}^\top$ 
             and the vortex. Here, we use $\text{\normalfont Ma}=0.1$.
             }\label{fig:thermal_vortex_init}
\end{figure}

We consider the propagation of a thermal vortex in a rectangular 
computational domain of size $L_x \times L_y = 150 \times 150$.
The initialization of the problem is described in normalized 
spatial coordinates $(\hat{x},\hat{y}) \in [-1,1]^2$:
\begin{equation}
   \hat{x} = \frac{2(x-1)}{L_x - 1  } -1, \quad \hat{y} = \frac{2(y-1)}{L_y - 1} -1.
\end{equation}
The initial center of the vortex is defined by:
\[
   (\hat{x}_0,\ \hat{y}_0) = \left( \tfrac{K}{L_x - 1}, \ 0 \right),
\]
where the parameter $K$ defines $x-$position of the vortex on the horizontal center line. 
The thermal vortex is formed by a perturbation of the temperature $T$ around $T_0$ within 
a circle at $(\hat{x}_0, \ 0)$ with radius $\hat{r}$ against a uniform background flow. 
We use the following initial macroscopic fields: 
\begin{flalign}
    &\rho(x, y) = \rho_0,  && \notag\\
    &\mathbf{u}(x,y) = \mathbf{u_0} + 
    \begin{cases}
    0 & \text{if} \   (\hat{x}- \hat{x}_0)^2 + \hat{y}^2 \geq \hat{r}^2 
    \\[1ex]
    \mathbf{v}(\hat{x}- \hat{x}_0,\hat{y})   & \text{otherwise},
    \end{cases} &&\\
    &T(x,y) = T_0 + 
    \begin{cases}
    0 & \text{if} \   (\hat{x}- \hat{x}_0)^2 + \hat{y}^2 \geq \hat{r}^2 \\
    \hat{y} \ \theta(\hat{x}- \hat{x}_0,\hat{y})   & \text{otherwise}, 
    \end{cases} \notag
&&
\end{flalign}
the vortex strength in terms of the initial perturbations is given in terms of a parameter $b$ as
\begin{flalign}
     &\mathbf{v}(x,y) 
        = 
        \frac{5 c_s \text{Ma}}{2} 2^{ -  \frac{x^2 + y^2}{b^2}}
        \begin{pmatrix}    y  \\[1ex] 
        -  x 
        \end{pmatrix}\!, \
     \notag   \\
    &   
        \theta(x,y) 
        = 
        \frac{5 c_s \text{Ma}}{2} 2^{ -  \frac{x^2 + y^2}{b^2}}. 
        \label{eq:initialvalues-vortex-example}
\end{flalign}
The setup is illustrated in Fig.~\ref{fig:thermal_vortex_init}.
We set $\text{Ma} = 0.1$, while the 
Eckert number corresponding to the simulation parameters is $\text{Ec} \approx 0.02$.

For the evaluation of the ZG BC, all boundary nodes are subjected to Eq.~\eqref{eq:zg-bc}.
The various characteristic based BCs summarized in Table~\ref{tab:CBCs} are applied  
at the right-hand side boundary, while the LODI scheme is used for the other straight boundaries.
Corner nodes are treated as discussed in Sec.~\ref{subsec:Realization}.

Contrary to the previous benchmark, this test exhibits fully two-dimensional dynamics. 
Therefore, it is more challenging with respect to the BC.
%

%

At the outlet (right-hand side boundary), we need to specify the 
wave amplitude $\mathcal{\bar{L}}_{x,3}$ (see Fig.~\ref{fig:orientation-L}).
The schemes with relaxation given by Eqns.~\eqref{eq:CBC-Relax} and \eqref{eq:LODI-Relax}
require the definition of the two relaxation parameters $\alpha$ and $\beta$. 
Following Ref.~\cite{yoo-ctan-2007}, we set $\alpha=\text{Ma}$,  $\beta=0,$ and  
$\mathcal{T}^{\infty}_{x,3}=0$. This choice of parameters implies that transversal 
waves vanish at the steady state. 
\begin{figure}[h]
  \centering
  \includegraphics[width=\linewidth]{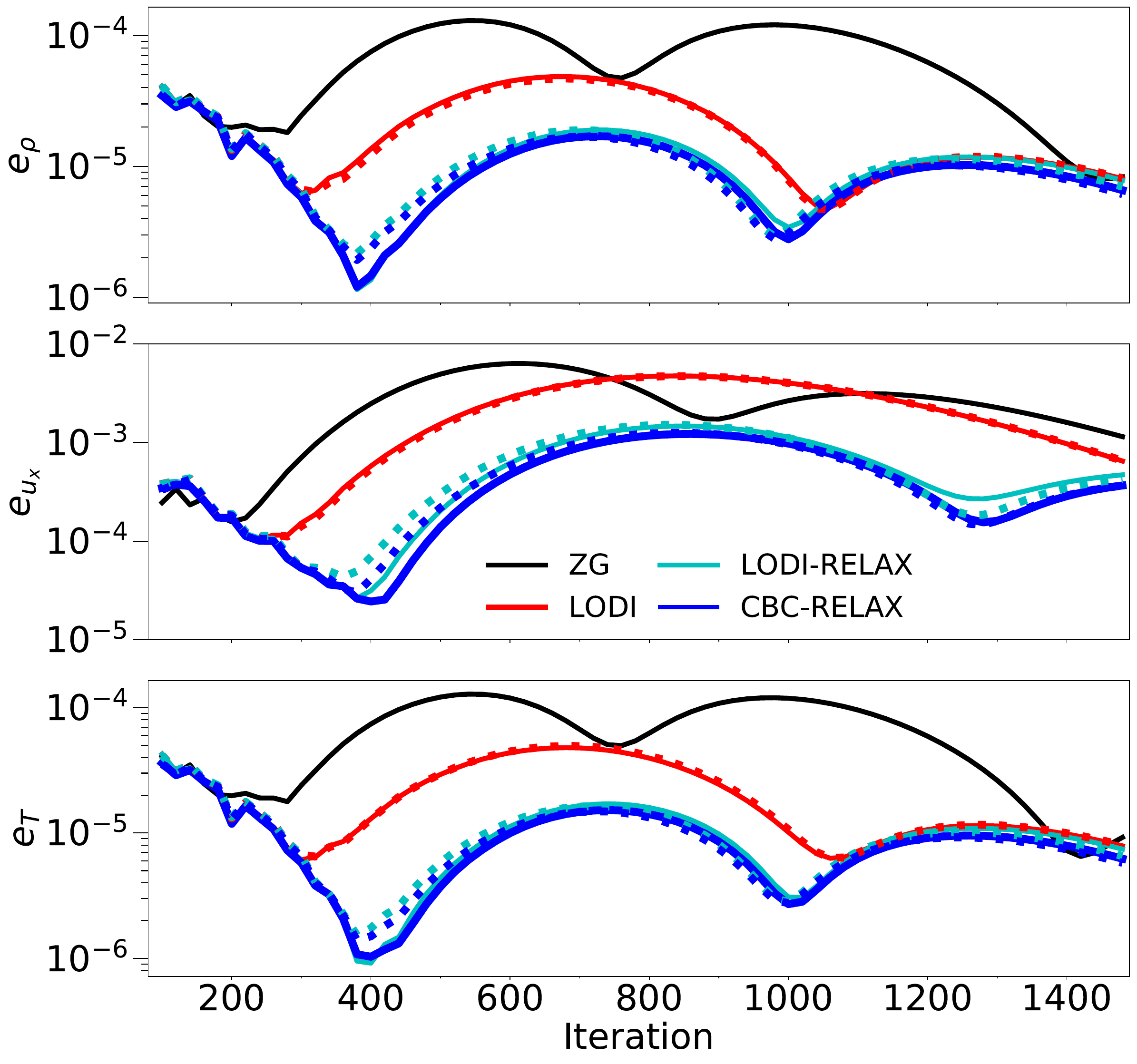}
  \caption{Evolution of relative $L^2$-errors for thermal vortex benchmark at $\nu=0.1$.
           We compare the error obtained employing different types of 
           characteristic BCs, combined with Dirichlet boundaries based on 
           the equilibrium BC (solid lines) and NEEP BC (dashed lines).
           }
  \label{fig:L2-Vortex}
\end{figure}

In the following, the equilibrium BC is used in conjunction with the characteristic BCs.
We start by considering a numerical viscosity of $\nu=0.1$.
The corresponding evolution of $e_Z$ for the various BC is shown in Fig.~\ref{fig:L2-Vortex}.  
It can be seen that in comparison with the LODI scheme, incorporating a relaxation of the 
transversal wave in the incoming wave amplitude reduces global errors by about one order 
of magnitude in the time range going between $400$ to $1000$ iterations.
Their similar error levels indicate that at least for this benchmark, 
encoding external information in the incoming wave amplitude is much more important than accounting 
for transversal information in the macroscopic system at the boundary.
We remark that, in general, it is not known how to choose the optimal values for the relaxation parameters.
For this specific benchmark, we have performed a parameter scan in $\alpha$ for values 
around $\alpha = \text{Ma}$ and found that the average values of $e_Z$ can be reduced 
by about $33$ percent for $\alpha = \frac{2}{3} \text{Ma}$.

The ZG BC gives similar results as the characteristic based BCs in the first $200$ iterations. 
Apart from the  iterations between $700$ and $1100$, where the value of $e_{u_x}$ is smaller 
than its LODI-counterpart, it gives the largest global errors among all the BC considered.

Similarly to what previously observed in Sec.~\ref{subsec:1dstep}, in Fig.~\ref{fig:L2-Vortex}
we show that the accuracy of the equilibrium BC is approximately the same as that of the NEEP.
For this reason, in what follows we will report results making use only of the equilibrium BC
to impose Dirichlet BCs.
\begin{figure}[htb]
  \centering
  \includegraphics[width=\linewidth]{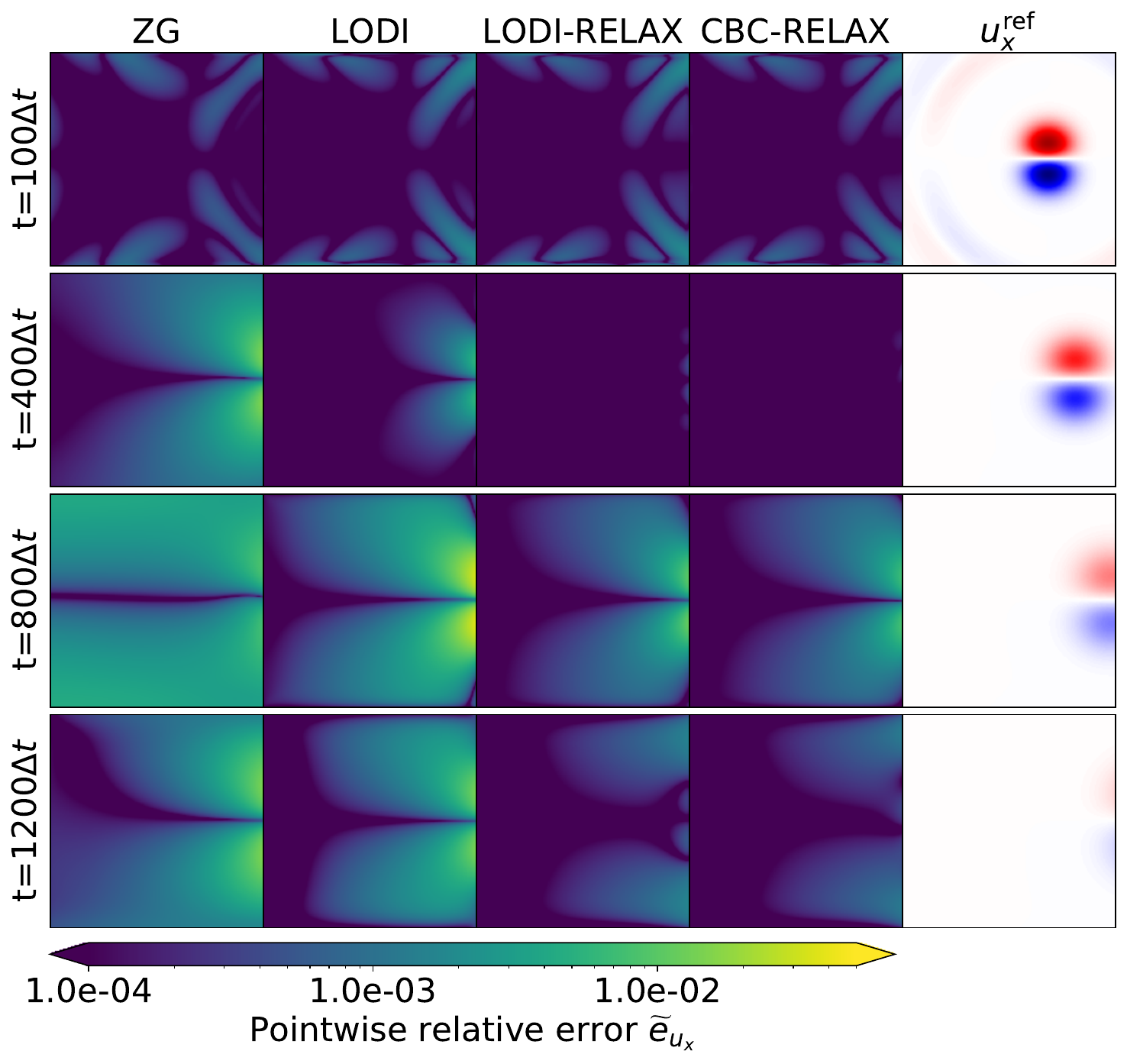}
  \caption{Heat maps of relative errors for the downstream velocity for various versions of the 
           BC conditions (vortex example). The right column give our reference solution. 
           In the rows, we have snap shots at $t=100\Delta t$ (initial phase), 
           $t=400\Delta t$ (the vortex has reached about the boundary), $t=800\Delta t$ 
           (center of the vortex is on the boundary), $t=1200\Delta t$ 
           (late phase, where the vortex has almost left the domain of interest).
           }\label{fig:Snap-ux-Vortex}
\end{figure}{}
\begin{figure}[htb]
  \centering
  \includegraphics[width=\linewidth]{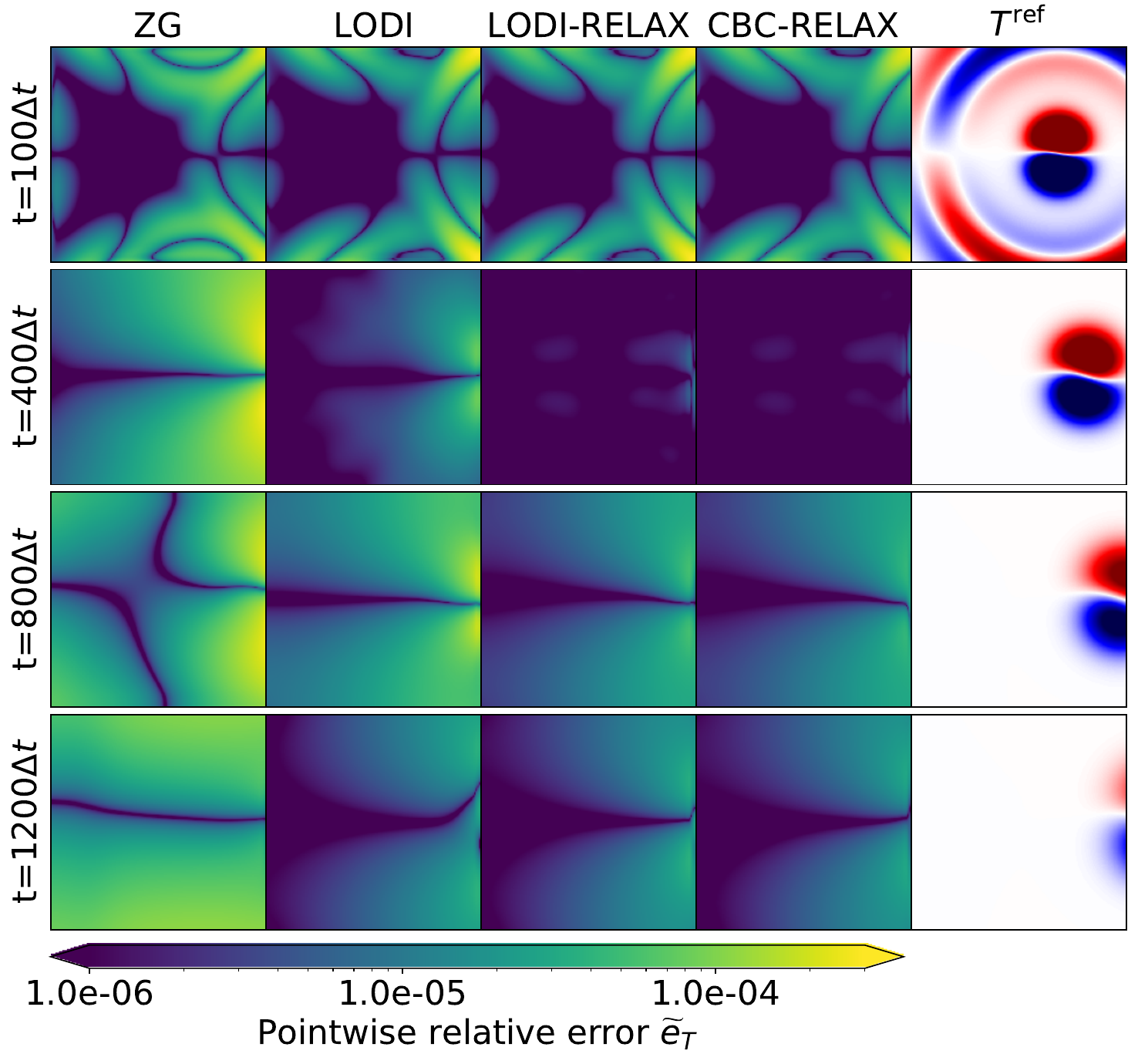}
  \caption{Heat maps of relative errors in temperature and reference solution on the right column 
           (vortex example). In the rows, we have snap shots at $t=100\Delta t$ (initial phase), 
           $t=400\Delta t$ (the vortex has reached about the boundary), $t=800\Delta t$ 
           (center of the vortex is on the boundary), $t=1200\Delta t$ 
           (late phase, where the vortex has almost left the domain of interest).
           }\label{fig:Snap-T-Vortex}
\end{figure}
In Figs.~\ref{fig:Snap-ux-Vortex} and \ref{fig:Snap-T-Vortex}, we provide snapshots of pointwise 
relative errors (with respect to the reference solution) for the downstream  velocity and temperature 
at selected time steps. 
For all the BC, small scale errors in both temperature and downstream velocity are observed at 
the boundaries  after $100$ iterations (first selected time step), as an initial spherical pressure 
pulse interacts with the boundaries. 
The vortex interaction with the boundary soon becomes the dominant source of error. 
As can be seen from the right-most columns in both figures, the vortex reaches the boundary after 
about $400$ iterations. From this point inwards, the relaxation approach gives a significant 
advantage over both the LODI and ZG BC. 
During the vortex-boundary interaction, here exemplified at $800$ and $1200$ iterations, the ZG is 
systematically outperformed. We observe that characteristic based schemes allow to better capture
the interaction with the boundary, and moreover the bulk region is significantly less polluted.

\paragraph{Treatment of viscous terms.\/}
We found the evaluation of viscous terms from Eqns~\eqref{eq:LaplaceU} 
and \eqref{eq:LaplaceT} crucial to ensure numerical stability over a broad range of 
numerical viscosity $\nu$.
To illustrate this, simulations at various values of $\nu$ have been conducted 
comparing with a CBC formulation where the Laplacian of temperature and velocity 
are approximated using second order finite differences instead of making use of Eqns.~\eqref{eq:LaplaceU} and \eqref{eq:LaplaceT}.
In particular, we use the backward formula for $j = M$
\begin{flalign}
     &\frac{\partial^2 \mathbb{U}_i(\mathbf{x}_{b,M},t)} {\partial^2 x} \approx
            2 \mathbb{U}_i( \mathbf{x}_{b,M},t) 
           -5 \mathbb{U}_i(\mathbf{x}_{b,M} - \mathbf{e}_x, t) && 
           \\ 
           &\hspace*{13ex} 
             +4  \mathbb{U}_i(\mathbf{x}_{b,M} - 2 \mathbf{e}_x, t)    
           - \mathbb{U}_i(\mathbf{x}_{b,M} - 3 \mathbf{e}_x, t)  &&  \notag
\end{flalign}
and central formula for the inner boundary nodes at $j =1,2,\dotsc, M-1$
\begin{flalign}
     &\frac{\partial^2 \mathbb{U}_i(\mathbf{x}_{b,j},t)} {\partial^2 x} 
       &&\\ 
       &\hspace*{2ex}\approx     
               \mathbb{U}_i(\mathbf{x}_{b,j} +  \mathbf{e}_x, t)
            -2   \mathbb{U}_i(\mathbf{x}_{b,j}                , t)  
            %
            +  \mathbb{U}_i(\mathbf{x}_{b,j} -  \mathbf{e}_x, t). \ \notag
         &&
\end{flalign}

The resulting scheme is referred to with the suffix -FD.
Note that this scheme can be used when working with stencils which do not allow to implement
high order quadrature.

\begin{figure*}[htb]
  \centering
  \includegraphics[width=\linewidth]{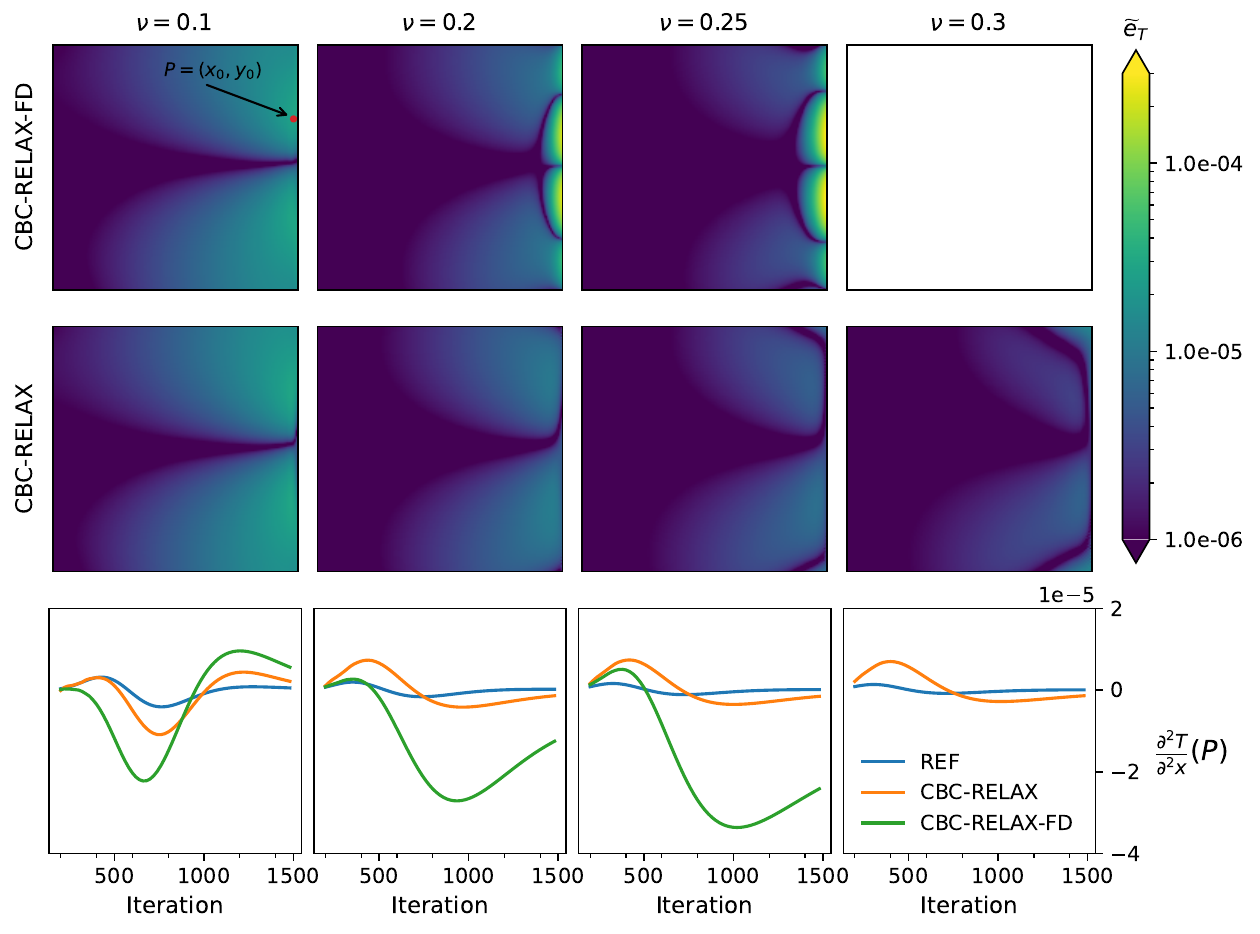}
    \caption{Upper and middle panels show heat maps of the relative errors in temperature $T$ 
             after $1200$ iterations at various numerical viscosity ranging from $\nu=0.1$ (first column) 
             to $\nu=0.3$ (right column). The lower panel shows the time evolution of the quantity 
             $\frac{\partial^2 T}{\partial x^2}$ computed at $P = (x_0, y_0)$ marked in red in 
             the top left panel. It can be seen that increasing the numerical viscosity, 
             the coupling between the macroscopic solver on the boundary and the LBM gives rise to an 
             instability when the Laplacian is evaluated directly at the macroscopic level rather than at the mesoscopic one;
             in particular, for $\nu=0.3$ (upper right panel) the simulation becomes numerically unstable.
             }
    \label{fig:L2-Vortex-VIS}
\end{figure*}

In the first two rows of Fig.~\ref{fig:L2-Vortex-VIS}, we show the 
heat maps of $\widetilde{e}_T$ after $1200$ iterations for several different values of the 
kinematic viscosity.
For $\nu =0.1$, we observe that the two methods yield very similar temperature fields.
In the last row of this figure, we report the time evolution of the Laplacian of $T$ evaluated
at point $P = (x_0, y_0)$ in proximity of the outlet (cf. top left panel in Fig.~\ref{fig:L2-Vortex-VIS}).
The mesoscopic evaluation of this quantity is closely following the evolution of ground truth,
whereas its finite difference counterpart exhibit larger and larger discrepancies, which become
more evident as the viscosity is increased, eventually leading to numerical instabilities at $\nu=0.3$.

\paragraph{Effect of the underlying velocity stencil.\/}
The characteristic BC developed in this work can be applied to any LBM stencil,
provided that the macroscopic target values returned by the artificial boundary
are supplied at the mesoscopic layer with a suitable Dirichlet BC.
As an example, in this section, we compare the results from simulations employing 
the D2Q37 velocity stencil. 
In Table~\ref{tab:CBCs-17-vs-37}, we report the arithmetic means, maximum value (over time) 
and the empirical standard deviation $s$ of error quantity $e_Z$ ($Z \in \{\rho,\, u_x,\, T\}$).
For a simple read out, errors are normalized with respect to the D2Q17 simulation using the ZG BC.

The overall behavior of errors is similar for both stencils. Switching from a ZG BC to LODI, 
the global errors in $\rho$ and $T$ decrease by roughly a factor of two, while the velocity fields 
are not significantly improved. However, tuning the relaxation parameters in the inward pointing 
wave amplitude allows to further decrease the errors in the velocity field. 
Switching the macroscopic equation that is solved on the boundary give further small 
improvements in accuracy.

The errors for $\rho$ and $T$ are approximately same for the two different stencils.

\begin{table*}
    \scalebox{0.95}{
        \begin{tabular}{ @{}|p{3.5mm}||*{1}{p{25.5mm}|p{11.5mm}}| }
            \hline
             & \multicolumn{2}{c|}{ZG\Tstrut}  \\
             \cline{2-3}
             & avg $\pm$ SEM\Tstrut & max  \\
             \hline\hline
             \scalebox{1.1}{$e_{\rho}^{17}$}\Tstrut & $6.4\! \cdot\!  10^{-5} \pm 4.8 \!\cdot\!  10^{-6}$ & $1.3 \!\cdot\!  10^{-4}$   \\[2pt] 
             \scalebox{1.1}{$e_{\rho}^{37}$}\Tstrut & $7.5\! \cdot\!  10^{-5} \pm 7.0 \!\cdot\!  10^{-6}$ & $1.6 \!\cdot\!  10^{-4}$   \\[2pt]
             \hline
             \scalebox{1.1}{$e_{u_x}^{17}$}\Tstrut & $2.5 \!\cdot\!  10^{-3} \pm  2.1 \!\cdot\!  10^{-4}$ & $6.3 \!\cdot \! 10^{-3}$  \\[2pt]
             \scalebox{1.1}{$e_{u_x}^{37}$}\Tstrut & $3.0 \!\cdot\!  10^{-3} \pm  3.0 \!\cdot\!  10^{-4}$ & $7.4 \!\cdot \! 10^{-3}$  \\[2pt]
             \hline
             \scalebox{1.1}{$e_{T}^{17}$}\Tstrut & $6.3 \!\cdot \! 10^{-5} \pm 4.8 \!\cdot\!  10^{-6}$ & $1.3 \!\cdot\!  10^{-4}$   \\[2pt] 
             \scalebox{1.1}{$e_{T}^{37}$}\Tstrut & $7.4 \!\cdot \! 10^{-5} \pm 6.9 \!\cdot\!  10^{-6}$ & $1.6 \!\cdot\!  10^{-4}$   \\[2pt] 
             \hline
         \end{tabular}
        \quad
        \begin{tabular}{|p{5mm}|*{3}{|p{5.5mm}|p{5.5mm}|}}
            \hline
             &  \multicolumn{2}{c||}{LODI\Tstrut}  & \multicolumn{2}{c||}{LODI-RELAX} & \multicolumn{2}{c|}{CBC-RELAX} \\
             \cline{2-7}
             &  avg\Tstrut  & max  & avg  & max  & avg  & max  \\
             \hline\hline
             \scalebox{1.1}{$e_{\rho}^{17}$}\Tstrut &  0.32   & 0.37  & 0.18  & 0.32  & 0.16  & 0.27  \\[2pt]
             \scalebox{1.1}{$e_{\rho}^{37}$}\Tstrut &  0.33   & 0.36  & 0.18  & 0.35  & 0.18  & 0.32  \\[2pt]
             \hline
             \scalebox{1.1}{$e_{u_x}^{17}$}\Tstrut &  0.82  & 0.75   & 0.22  & 0.23   & 0.18  & 0.19  \\[2pt]
             \scalebox{1.1}{$e_{u_x}^{37}$}\Tstrut &  0.78  & 0.77   & 0.22  & 0.22   & 0.18  & 0.18  \\[2pt]
             \hline
             \scalebox{1.1}{$e_{T}^{17}$}\Tstrut &  0.32  & 0.37   & 0.17  & 0.33   & 0.15  & 0.28   \\[2pt] 
             \scalebox{1.1}{$e_{T}^{37}$}\Tstrut &  0.34 & 0.37   & 0.18  & 0.36   & 0.17  & 0.32   \\
             \hline
         \end{tabular} 
     }
     \caption{Comparison of the obtained accuracy the vortex benchmark using D2Q17 and D2Q37 
              velocity stencils: We compare arithmetic mean of the sampled global errors $e_Z$, 
              the corresponding standard error of the mean (SEM) and their maximum value (over time). 
              Results are normalized with respect to the values obtained for the ZG BC and the 
              respective stencil (left table).  
              The SEM has been omitted from the right-hand side table as it scales similarly 
              to the average error.
              }\label{tab:CBCs-17-vs-37}
\end{table*}

\subsection{Angular Wave}\label{subsec:ang-wave}
In this section, we consider an impinging plane wave that approaches the boundary at an angle $\phi$ 
with respect to the vertical line ($x=x_b$, left boundary); i.e., $\phi=0$ states that the plane wave 
propagates in the direction normal to the boundary.
In this setup, as the value of $\phi$ is increased, the transversal contributions become more and 
more important, in turn departing from the LODI approximation.

We evaluate the BCs ability to absorb outgoing information by computing a reflection coefficient. 
To this end, we calculate wave amplitudes before and after the interaction 
with the boundary takes place (from simulation data) and compute their ratio along the horizontal midplane.
\begin{figure}[h]
  \centering
  \includegraphics[width=\linewidth]{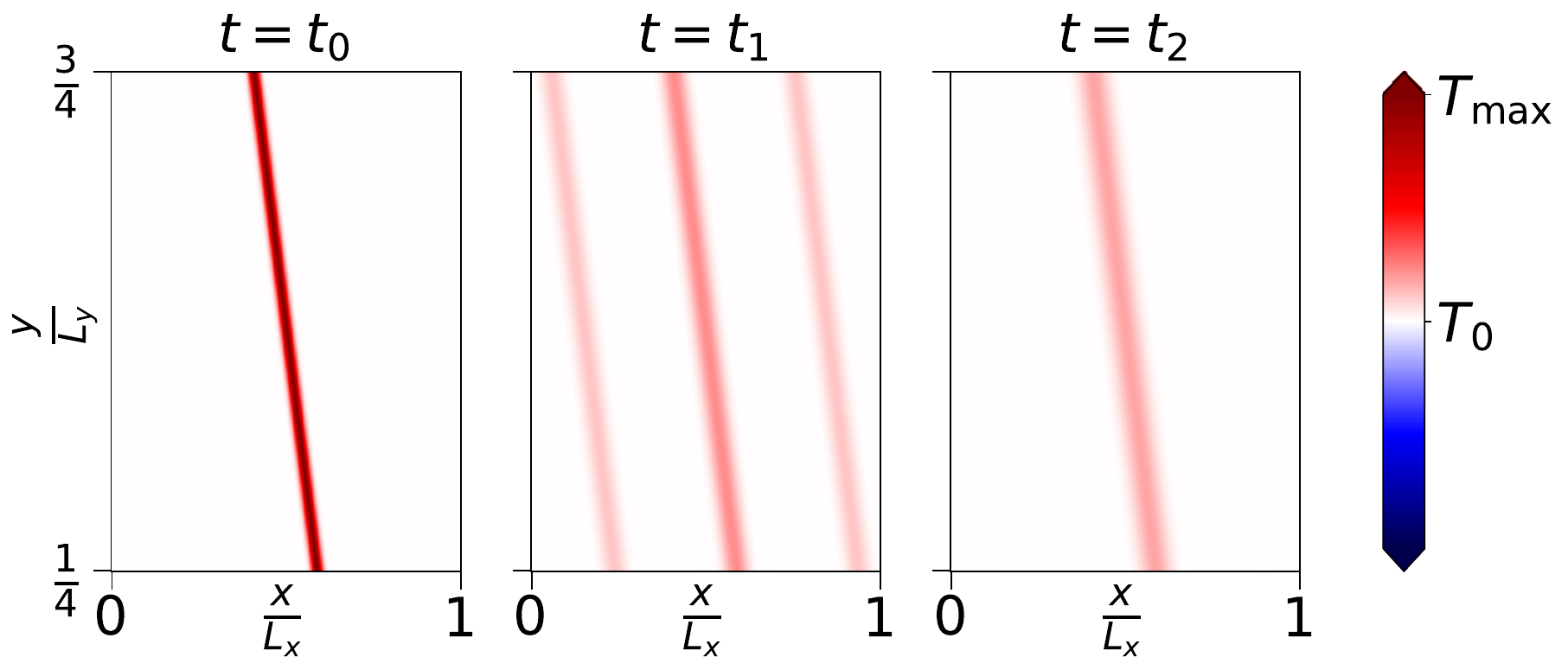}
  \caption{Temperature fields obtained using the LODI BC at various time steps. 
           The left panel depicts the initial state at time $t_0$ with initial conditions given by
           $\rho_0 = T_0 = 1,\; \mathbf{u_0} = \vek{0,&0}^\top$ and $s=\frac{1}{50}$. 
           The times $t_1$ and $t_2$ correspond to before and after waves interacted with the 
           left hand side boundary at height $y=\frac{L_y}{2}$ respectively. 
           }\label{fig:ang-wave-timesnaps}
\end{figure}

The initial setup at time $t_0$ reads
\begin{align}
    \rho(x,y) &= \rho_0, \
    \mathbf{u}(x,y) = \vek{0,&0}^\top, \notag \\
    T(x,y) &= T_0 + \frac{1}{10} \exp\left( \frac{-\hat{x}(x,y)^2}{2s^2}\right),
\end{align}
where the shifted coordinates 
\begin{equation}
   \hat{x}(x,y) = \vek{\cos( \phi \frac{\pi}{180}) \\[1ex] \sin( \phi \frac{\pi}{180})} \cdot \vek{x \\[1ex] y }
\end{equation}
are used.
As the simulation develops, two pressure (temperature) pulses travel along the positive 
and negative $x-$ direction respectively while the initial temperature spike dissipates.
The upper and lower boundaries are periodic, while artificial boundaries are set at the left 
and right-hand side of the domain.
We chose a numerical viscosity of $\nu=0.1$ and conducted simulations on a 
$L_x \times L_y = 200 \times 700$ grid. The domain is chosen to be sufficiently large to ensure 
that the measurements at the horizontal centerline are not polluted by artifacts stemming 
from the periodicity of the upper and lower boundaries.
We perform simulations at various angles $\phi$, tracking wave amplitudes $I_Z$ for the
macroscopic quantities $Z \in \{ \rho, u_x, T \}$ along the horizontal slice $y= \frac{L_y}{2}$. 
The measurements are taken at a time $t_1$ and $t_2$, respectively shortly before 
and right after the interaction with the boundaries (cf. Fig.~\ref{fig:ang-wave-timesnaps}).
The reflection coefficient is then computed as 
\begin{equation}
    R_Z = \frac{I_Z(t_2)}{I_Z(t_1)}.
\end{equation}
We use $R_Z$ to quantify errors introduced by the boundary treatment. Note that in the ideal 
case for which outgoing waves exit the computational domain without introducing any reflection 
$R_Z$ would be exactly zero.
Snapshots of the temperature profiles at times $t_0,t_1,t_2$ are shown in Fig.~\ref{fig:ang-wave-timesnaps} 
showing the initial state and the state before and after the boundary interaction of the angular wave.
\begin{figure}[htb]
    \centering
    \includegraphics[width=\linewidth]{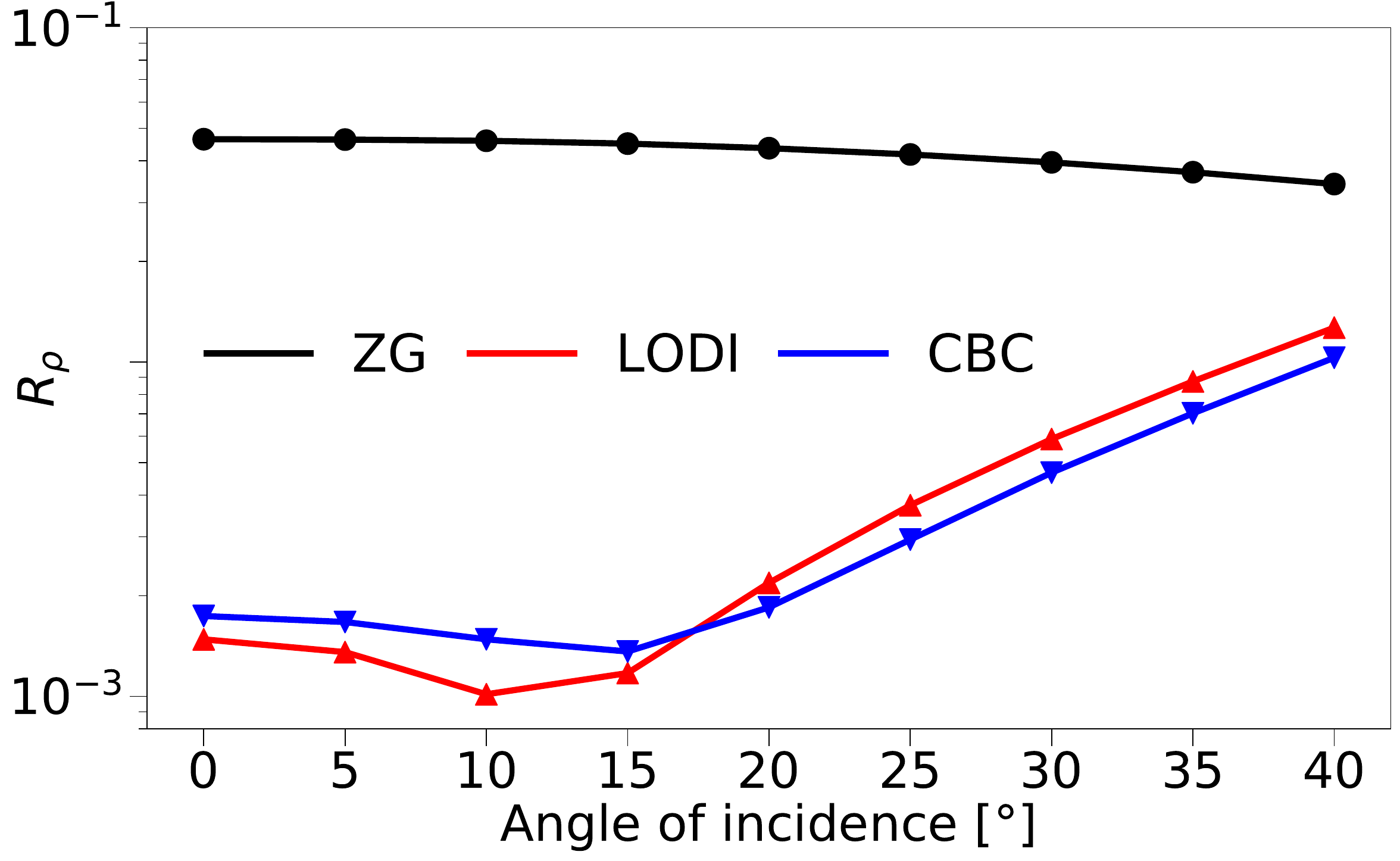}
    \caption{Angular dependency of the reflection coefficient $R_{\rho}$ along the horizontal 
             slice $y= \frac{L_y}{2}$, on a $L_x \times L_y = 200 \times 700$ grid.        
                }
            \label{fig:angular_wave_ang_dependency}
\end{figure}
In this benchmark, the characteristic BCs are considered in conjunction with the mesoscopic equilibrium BC \eqref{eq:feq-3}.
In Fig.~\ref{fig:angular_wave_ang_dependency}, we report the reflection coefficient $R_{\rho}$
over a range of values for the angle $\phi$. The reflection coefficients for temperature and 
streamwise velocity gives very similar results and are hence omitted here.
We observe that the ZG BC gives rise to reflected waves with about four percent of the impinging waves amplitude. 
For small angles $\phi < 15^{\circ}$, the usage of the (perfectly non-reflecting) CBC reduces reflections 
by roughly one order of magnitude. As $\phi$ is increased further, we observe a growth in
reflections caused by the characteristic based schemes. This can be attributed to two factors.
First, as already mentioned, for non-zero values of $\phi$ the dynamic starts departing 
from the LODI approximation. Second, waves impinging at large values of $\phi$
interact with the boundary for longer time with respect to waves at smaller angles,
putting further stress on the BC.
Nevertheless, the advantage of CBC over the ZG BC is still retained 
at a $40^{\circ}$ angle, where reflection coefficient is reduced by a factor of about three.

As already discussed in Sec.~\ref{subsec:1dstep}, the reincorporation of transversal terms in the 
CBC scheme gives rise to small oscillations close to the outlet. 
This leads to slightly higher reflection coefficients at angles $\phi <25^{\circ}$ 
when compared to the LODI scheme.
However, at angles $\phi >25^{\circ}$, the  error in discarding transversal information  
becomes dominant and the LODI scheme gives rise to larger reflection coefficients than the CBC.

\subsection{Corners and multi-speed models}\label{subsec:corners-multi-speed}
In previous sections, we have discussed implementation and numerical results for characteristic
based BC applied to multi-speed thermal LBM.
These models require the application of BC to several layers of nodes, which can lead to small
oscillations at the interface between the boundary and the bulk domain (see again 
Fig.~\ref{fig:thermal-step-rel-err-pointwise2}), although such spurious effects can be mitigated
resorting to a mesoscopic evaluation of the partial derivatives for velocity and temperature field
(Fig.~\ref{fig:L2-Vortex-VIS}).
The complexities associated to handling multiple boundary layers are emphasized when dealing with
corner nodes (see Fig.~\ref{fig2:Layer-Multispeed-BC-v2}). While there exist compatibility 
conditions \cite{lodato-jocp-2008} to be posed at overlapping boundaries, their application to corners
formed by several nodes in a multi-speed setting is not immediately obvious.
For this reason, we have relied on a simple LODI approximation for the treatment of corners 
of the computational domain. 
In order to assess the impact of corners in simulation results, in this section, we consider 
simulations for the propagation of a iso-thermal vortex. 
The numerical setup is exactly the same as for the thermal flow discussed in Sec.~\ref{subsec:vortex}
with the only difference that we start now from a uniform temperature profile, which is kept constant
during the time evolution of the flow.
This allows to compare the results of simulations from multi-speed models with those given by 
the single-speed D2Q9 model. Details on all the stencils used are given in Appendix B.
Since the D2Q9 does not correctly capture the heat-flux, we employ here the CBC-RELAX-FD 
scheme discussed in Sec.~\ref{subsec:vortex}. 
At the boundary we use the following parameters: $\alpha=\text{Ma },\beta=0,$ and  $\mathcal{T}^{\infty}_{x,3}=0$. 
The corners treatment for the D2Q9 is analogous to the multi-speed case (c.f. Fig.~\ref{fig2:Layer-Multispeed-BC-v2}),
albeit simpler since target quantities $\mathbb{U}$ needs to be calculated for one single corner node.

\begin{figure}[htb]
  \centering
  \includegraphics[width=\linewidth]{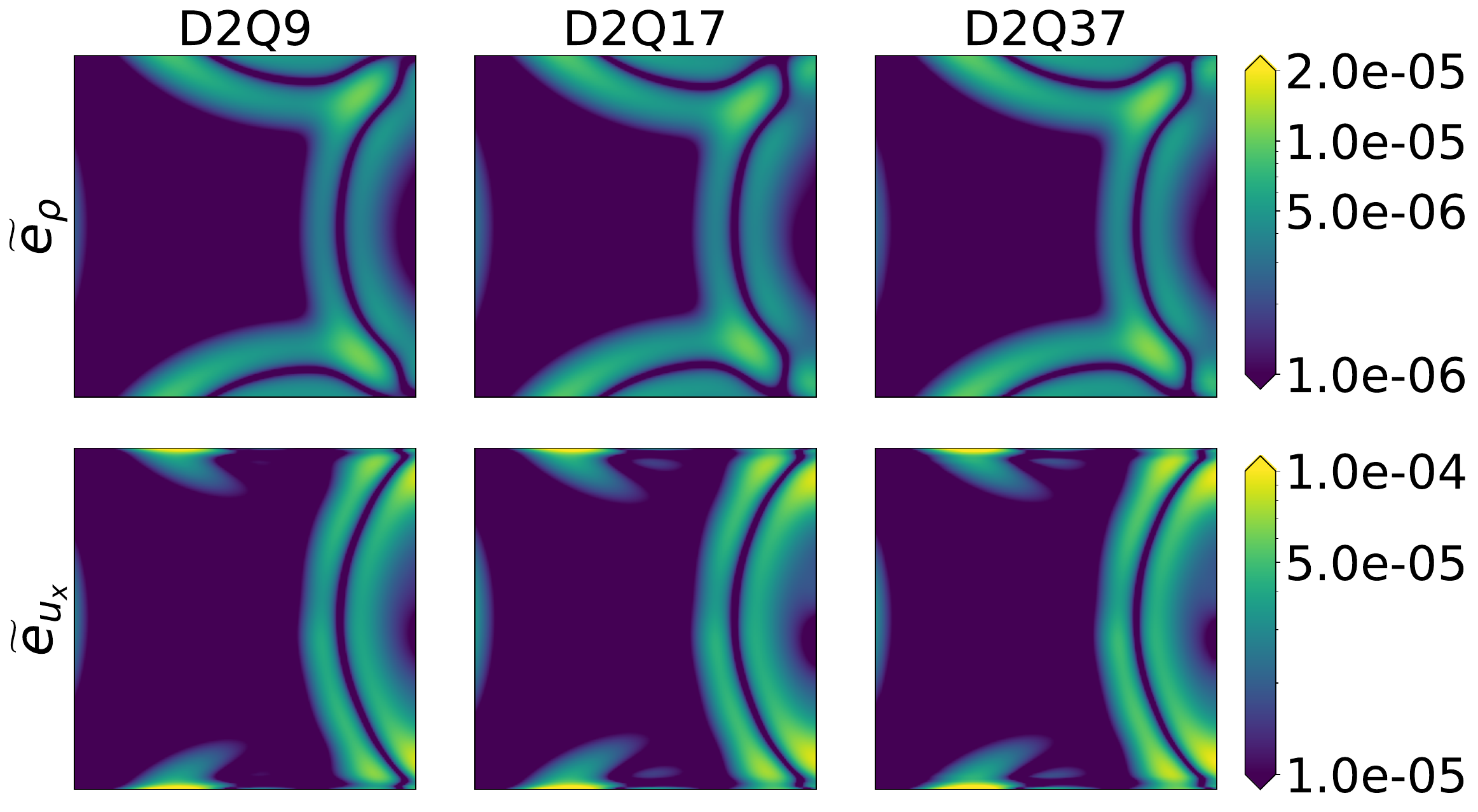}
    \caption{Comparison of pointwise relative errors obtained in the simulation of an 
             iso-thermal vortex using the D2Q9, D2Q17 and D2Q37 velocity stencil.
             }
    \label{fig:Vortex-Corner-Stencil-Cmp}
\end{figure}

In Fig.~\ref{fig:Vortex-Corner-Stencil-Cmp}, we show heat maps for both quantities $\widetilde{e}_{\rho}$ and 
$\widetilde{e}_{u_x}$. 
In order to allow for a direct comparison between the different numerical scheme, we show results at 
the re-scaled time $t^* = \lfloor 150 c_s^{17} / c_s^{q} \rfloor \quad q \in \{ 9, 17, 37 \}$, where
$c_s^{q}$ is the speed of sound in the lattice for the different stencils (see Appendix B).
Such a value is chosen in order to analyze a time frame where reflections caused at the outlet 
and lateral boundaries are interacting with each other close to the corners of the computational domain, which is
putting under stress the corner boundary.  
From the results, we can observe that the dynamic obtained with the D2Q9 stencil closely resembles
that provided by multi-speed stencils, with minor differences observed 
close to the right-hand side corners. Nevertheless, the pointwise errors in the proximity of the corners 
are of the same order of magnitude as in the remaining points close to the outlet.
Therefore, we can conclude that the potentially extra source of inaccuracy given by the 
corner treatment in multi-speed stencils is negligible, at least for the benchmark here considered.

\section{Conclusion}\label{sec:conclusions}
%
In this work, we have presented a non-reflecting BC for thermal 
LBM, applying the concept of characteristic boundary conditions 
to flows governed by the Navier-Stokes-Fourier system.
The procedure allows to compute outgoing wave amplitude variations
from a hyperbolic system using the LODI approximation, and aims
at modulating the amplitude of incoming waves in order to minimize
their impact on the bulk dynamics. 
By constraining the amplitude of incoming waves (annihilation, 
relaxation towards target values), it is possible to compute
the macroscopic fields at the boundary layer, which are then 
translated at the mesoscopic level into lattice populations.

While the procedure is general and can be applied to any of the
different possible approaches for the simulation of thermal flows
in LBM, we have focused our analysis on high order models based 
on multi-speed stencils.
We have shown that this approach offers the advantage that 
the evaluation of the Laplacian for temperature and velocity can
be established exploiting the exact calculation of the higher order
moments of the particle distribution function, hence on the same 
mesoscopic footstep as LBM. 
Our numerical results highlight that this approach leads to
more accurate results as well as enhanced stability at increasingly
large values of the kinematic viscosity.
Moreover, we have compared several CBC schemes against 
a zero-gradient BC, a commonly adopted strategy for the implementation
of artificial boundaries in literature.
Table~\ref{tab:summary} provides a summary of our findings, showing 
that for flows with a strongly one-dimensional propagation direction, i.e., where
the LODI approximation fully applies, the CBC outperforms 
the ZG BC by 2 to 3 order of magnitudes. This gap is reduced to about
one order of magnitude when investigating flows with 
a significant transversal component. In the second case, further
improvement can be obtained by relaxing the macroscopic fields to
desired target values, although this generally requires tuning of
extra parameters, for which the optimal value is not known a priori.

As a closing remark, we shall stress that our current implementation 
for translating macroscopic quantities into mesoscopic 
lattice populations is generally based on a coarse grained
approach, aiming at simplifying the treatment 
of multiple adjacent grid points required by multi-speed LBM stencils.
The main approximations used in this work are i) the full replacement
of information at the boundary cells using an equilibrium BC
and ii) the definition of macroscopic quantities at corner nodes
by conducting the characteristic analysis with respect of 
one single neighboring node in the bulk along the diagonal direction.

We have shown that for the benchmarks reported in our analysis, 
the equilibrium BC can yield similar level of accuracy 
as for cases where the non-equilibrium part of the distribution 
was extrapolated from neighboring bulk cells.
However, we expect that more advanced Dirichlet BC schemes,
such as Non-Equilibrium Bounce-Back BC \cite{klass-jocs-2021} 
and regularized BC schemes~\cite{malaspinas-cf-2011,frapolli-pre-2014},
will help increasing accuracy and stability in more
involved kinematic regimes
e.g. at large values for the Reynolds and Rayleigh number.
We also plan to investigate the effect of making use of 
partial information already available (post-streaming)
at the boundary cells.
Concerning corner nodes, we plan to define 
suitable compatibility conditions~\cite{lodato-jocp-2008}
allowing to perform the characteristic analysis by combining 
information from both the $x$- and $y$-direction, even for cases
where corners are placed at the interface between different
types of BC.

Finally, we also plan to evaluate the possibility
of combining CBC with flux-splitting methods \cite{liou-jocp-1993},
and the perfectly matched layer approach \cite{najafi-cf-2012}.

\begin{table}
\begin{tabular}{|p{6mm}|*{2}{|p{19mm}|p{6mm}|}}
     \hline
     &  \multicolumn{2}{c||}{Temperature step: CBC} & \multicolumn{2}{c|}{Vortex: CBC-RELAX}   \\
     \cline{2-5}
     & avg & max   & avg & max  \\
     \hline\hline
     \scalebox{1.1}{$e_{\rho}$} & 0.007  & 0.009 & 0.16  & 0.27 \\[2pt]
     \hline
     \scalebox{1.1}{$e_{u_x}$} & 0.001  & 0.002  & 0.18  & 0.19 \\[2pt]
     \hline
     \scalebox{1.1}{$e_{T}$} & 0.006  & 0.009 & 0.15  & 0.28 \\
     \hline
 \end{tabular}
 \caption{Summary of the gains in accuracy normalize with respect to the global relative errors 
          $e_Z$ obtained using the D2Q17 stencil and the ZG BC.
          Shown are the arithmetic mean of the sampled global errors $e_Z$ 
          and their maximum value (over time). 
          The SEM has been omitted as it scales very similar to the average error.
          }\label{tab:summary}
\end{table}

\section*{Acknowledgments}
%
The authors would like to thank Matthias Ehrhardt (University of Wuppertal) for useful discussions.
The authors are grateful to an anonymous referee of "Computers and Mathematics with Applications"
for the critical reading of the manuscript and for many useful comments.

\section*{Appendix}\label{sec:appendix}
 
\subsection*{A. CBC and LODI for Navier-Stokes-Fourier in $d=3$ spatial dimensions}

In this appending section, we provide the ingredients necessary to extend the implementation of the 
characteristic based BC for the Navier-Stokes-Fourier in $d=3$ spatial dimensions.
We follow the same procedure discussed in the main text, and consider as an example 
the case of a right-hand side boundary.
Here, the macroscopic velocity is given as $\mathbf{u}=(u_x,u_y,u_z)^\top$ 
and the transversal directions are $\mathbf{u}_t=(u_y,u_z)^\top$ (for a boundary $x=x_b$ constant).
Furthermore, we denote the  spatial gradient in the transversal directions as 
$\nabla_t=(\frac{\partial}{\partial y}, \frac{\partial}{\partial z})^\top$.

The specific heat quantities read 
\begin{equation}
    c_v=\frac{d}{2}=\frac{3}{2}, c_p=\frac{d}{2}+1=\frac{5}{2} 
\end{equation}
with ratio $\gamma=\frac{c_p}{c_v}=\frac{5}{3}$.

Following the same matrix representation from \eqref{eq:NSF-CHAR}, we have here
\begin{flalign*}
    &A = 
    \begin{pmatrix}
        u_x                  & \rho & 0   & 0 & 0    \\[0.5ex]
       \frac{ \tilde{T} }{\rho} & u_x  & 0 & 0  & c_s^2 \\[0.5ex]
        0                    & 0    & u_x & 0   & 0    \\[0.5ex]
        0                    & 0    & 0   & u_x & 0    \\[0.5ex]
        0                    & \frac{\tilde T}{c_s^2 c_v}    & 0   & 0 &   u_x
    \end{pmatrix}, &&
   \\
	&
    \mathbb{T} =  \vek{ -   \frac{\partial (\rho u_y)}{\partial y} - \frac{\partial (\rho u_z)}{\partial z} \\[0.5ex]
         - \mathbf{u}_t \cdot \left( \nabla_t  \ u_x \right) \\[0.5ex]
         - \frac{1}{\rho} \frac{\partial (\rho \tilde{T})}{\partial y}  - \mathbf{u}_t \cdot \left( \nabla_t  \ u_y \right) \\[0.5ex]
         - \frac{1}{\rho} \frac{\partial (\rho \tilde{T})}{\partial z} - \mathbf{u}_t \cdot \left( \nabla_t  \ u_z \right) \\[0.5ex]
         -  \frac{\tilde T}{c_s^2 c_v} \nabla_t \cdot \mathbf{u}_t -  \frac{1}{c_s^2} 
         \mathbf{u}_t \cdot  \nabla_t \tilde T
     }&&
\end{flalign*}
and 
\begin{flalign*}
    &\mathbb{V} =  \vek{ 0 \\[0.5ex]
    \nu \left( \Delta u_x + \frac{1}{3} \frac{\partial}{\partial x} \mathrm{div} \ \mathbf{u} \right) \\[0.5ex]  
    \nu \left( \Delta u_y + \frac{1}{3} \frac{\partial}{\partial y} \mathrm{div} \ \mathbf{u} \right)\\[0.5ex]
    \nu \left( \Delta u_z + \frac{1}{3} \frac{\partial}{\partial z} \mathrm{div} \ \mathbf{u} \right) \\[0.5ex]  
    \frac{\nu \gamma}{ \text{Pr} \  c_s^2} \Delta \tilde T + \frac{\nu}{c_v c_s^2} \left( V_1 + \frac{2}{3} V_2 \right)
    }, &&
\end{flalign*}
where \begin{flalign*}
    &  V_1    = \left( \frac{\partial u_x}{\partial y} + \frac{\partial u_y}{\partial x} \right)^{\!\! 2} 
        + \left( \frac{\partial u_x}{\partial z} + \frac{\partial u_z}{\partial x} \right)^{\!\! 2}
        + \left( \frac{\partial u_y}{\partial z} + \frac{\partial u_z}{\partial y} \right)^{\!\! 2}  && \\
        & V_2 =  \left( \frac{\partial u_x}{\partial x} - \frac{\partial u_y}{\partial y} \right)^{\!\! 2}
        + \left( \frac{\partial u_x}{\partial x} - \frac{\partial u_z}{\partial z} \right)^{\!\! 2} 
        +\left( \frac{\partial u_y}{\partial y} - \frac{\partial u_z}{\partial z} \right)^{\!\! 2}.
\end{flalign*}

The diagonalization of $A$ gives $A = S^{-1} \Lambda S$ 
with 
\begin{align*}
	\Lambda = &
	\textrm{diag}\left(u_x, u_x, u_x, u_x - \sqrt{ \gamma \tilde{T}}, u_x + \sqrt{\gamma \tilde{T}}\right),  \\
    S = &
    \begin{pmatrix}
    -\frac{2 \tilde T}{5 \rho c_s^2}  & 0                                   & 0  & 0 & \frac{3}{5} \\[0.7ex]
    0                                 & 0                                   & 0  & 1 & 0           \\[0.7ex]
    0                                 & 0                                   & 1  & 0 & 0           \\[0.7ex]
    \frac{\tilde T}{5 \rho c_s^2}     & -\sqrt{\frac{\tilde T}{15 c_s^4}}   & 0  & 0 & \frac{1}{5} \\[0.7ex]
    \frac{\tilde T}{5 \rho c_s^2}     & \sqrt{\frac{\tilde T}{15 c_s^4}}    & 0  & 0 & \frac{1}{5} \\[0.7ex]
    \end{pmatrix},
 \\
    S^{-1} = &
    \begin{pmatrix}
    -\frac{\rho c_s^2}{\tilde T} &  0  &  0 & \frac{3 \rho c_s^2}{2 \tilde T}      & \frac{3 \rho c_s^2}{2 \tilde T}   \\[0.7ex]
    0                            &  0  &  0 & - \sqrt{\frac{15 c_s^4}{4 \tilde T}} & \sqrt{\frac{15 c_s^4}{4 \tilde T}}   \\[0.7ex]
    0                            &  0  &  1 & 0                                    & 0 \\[0.7ex]
    0                            &  1  &  0 & 0                                    & 0 \\[0.7ex]
    1                            &  0  &  0 & 1                                    & 1
    \end{pmatrix}. 
\end{align*}

The macroscopic fields at the boundary can be then calculated using the CBC approach by solving 
Eq.~\eqref{eq:thermal_CBC_macro_evo}, or Eq.~\eqref{eq:thermal_LODI_macro_evo} for LODI.
%

\subsection*{B. Data on velocity stencils}

In this appendix section we provide details on the velocity stencils, quadrature weights and speed 
of sound in the lattice $c_s$ used to implement the high order LBM models used in simulations 
in the main text. For the sake of completeness, 
we also provide details for the single-speed D2Q9 lattice used in Sec.~\ref{subsec:corners-multi-speed}.

The information for the D2Q9 is provided in Tab.~\ref{tab:d2q9}. Data for the multi-speed  
D2Q17 is given in Tab.~\ref{tab:d2q17}, while in Tab.~\ref{tab:d2q37} we give the data for the D2Q37. 
In all tables, each row is associated to a different fully symmetric set of 
discrete velocities, where the following notation is implied: e.g. 
$(1,1) = \{ (-1,-1), (-1,1), (1,-1), (1,1) \} $.

In order to implement CBC for thermal LBM simulation in $d = 3$ spatial dimensions one can make use
e.g. of the D3Q39 and the D3Q103 stencils, for which details on quadrature data can be found in
Ref.~\cite{shan-jocs-2016}.

\begin{table}[bhtb]
\centering
\begin{tabular}{|c|c|c|}%
   \hline
   indices      $i$       & velocities $c_i$  & weights $\omega_i$ \\
   \hline
                 $1$      & $(0,0)$           & $4/9$    \\
                 $2-5$    & $(1,0)$           & $1/9$    \\
                 $6-9$    & $(1,1)$           & $1/36$    \\                
   \hline
\end{tabular}
\caption{Discrete velocity set for the D2Q9 stencil. 
         The stencil speed of sound is $c_s = \frac{1}{\sqrt{3}}$. 
        }\label{tab:d2q9}
\begin{tabular}{|c|c|c|}%
   \hline
   indices      $i$       & velocities $c_i$  & weights $\omega_i$ \\
   \hline
                 $1$      & $(0,0)$           & $\frac{575+193 \sqrt{193}}{8100}$    \\
                 $2-5$    & $(1,0)$           & $\frac{3355-91 \sqrt{193}}{18000}$    \\
                 $6-9$    & $(1,1)$           & $\frac{655+17 \sqrt{193}}{27000}$    \\
                 $10-13$  & $(2,2)$           & $\frac{685-49 \sqrt{193}}{54000}$    \\
                 $14-17$  & $(3,0)$           & $\frac{1445-101 \sqrt{193}}{162000}$    \\
   \hline
\end{tabular}
\caption{Discrete velocity set for the D2Q17 stencil. 
         The stencil speed of sound is $c_s =\sqrt{ \frac{5 (25 + \sqrt{193})}{72}}.$
        }\label{tab:d2q17}
\begin{tabular}{|c|c|c|}%
   \hline
   indices      $i$       & velocities $c_i$  & weights $\omega_i$ \\
   \hline
                 $1$      & $(0,0)$           & 0.23315066913235    \\
                 $2-5$    & $(1,0)$           & 0.10730609154221    \\
                 $6-9$    & $(1,1)$           & 0.05766785988879    \\
                 $10-13$  & $(2,0)$           & 0.01420821615845    \\
                 $14-21$  & $(2,1)$           & 0.00535304900051    \\
                 $22-25$  & $(2,2)$           & 0.00101193759267    \\
                 $26-29$  & $(3,0)$           & 0.00024530102775    \\
                 $30-37$  & $(3,1)$           & 0.00028341425299    \\
   \hline
\end{tabular}
\caption{Discrete velocity set for the D2Q37 stencil. 
         The stencil speed of sound is $c_s =\frac{1}{6} \sqrt{49- \frac{119+\left(469+252\sqrt{30}\right)^{\frac{2}{3}}}{\left(469+252\sqrt{30}\right)^{\frac{1}{3}}}}$. 
        }\label{tab:d2q37}
\end{table}


\bibliographystyle{elsarticle-num}




\end{document}